\documentclass[10pt,letterpaper]{article}
\usepackage[top=0.85in,left=1.75in,footskip=0.75in,marginparwidth=2in]{geometry}

\usepackage[utf8]{inputenc}

\usepackage[
backend=biber,
style=numeric,
sorting=none
]{biblatex}
\usepackage{nameref,hyperref}

\usepackage[right]{lineno}

\usepackage{microtype}
\DisableLigatures[f]{encoding = *, family = * }

\usepackage{changepage}

\usepackage[aboveskip=1pt,labelfont=bf,labelsep=period,singlelinecheck=off]{caption}

\makeatletter
\renewcommand{\@biblabel}[1]{\quad#1.}
\makeatother

\usepackage{lastpage,fancyhdr,graphicx}
\usepackage{epstopdf}
\fancyheadoffset[L]{2.25in}
\fancyfootoffset[L]{2.25in}

\usepackage{color}

\definecolor{Gray}{gray}{.25}

\usepackage{graphicx}

\usepackage{sidecap}

\usepackage{wrapfig}
\usepackage[pscoord]{eso-pic}
\usepackage[fulladjust]{marginnote}
\reversemarginpar

\usepackage{amsmath}
\usepackage{booktabs}
\usepackage{longtable}
\usepackage{multirow}
\usepackage[capitalise]{cleveref}
\usepackage{tabularx}
\usepackage{threeparttable}
\usepackage{ragged2e}
\usepackage{enumitem}
\usepackage{wrapfig}

\crefformat{equation}{(Eq.~#2#1#3)}
\crefformat{figure}{(Fig.~#2#1#3)}
\crefformat{table}{(Tab.~#2#1#3)}
\crefname{equation}{Eq.}{equations}

\newcolumntype{b}{X}
\newcolumntype{s}{>{\hsize=.25\hsize}X}

\newcommand\numberthis{\addtocounter{equation}{1}\tag{\theequation}}

\begin{document}
\vspace*{0.35in}

\begin{flushleft}
{\Large
\textbf\newline{Integrating time-resolved \textit{nrf2} gene-expression data into a full GUTS model as a proxy for toxicodynamic damage in zebrafish embryo}
}
\newline
\\
{
    \bf
    Florian Schunck\textsuperscript{1,*},
    Bernhard Kodritsch\textsuperscript{2},
    Wibke Busch\textsuperscript{2},
    Martin Krauss\textsuperscript{2},
    Andreas Focks\textsuperscript{1}
}
\\
\bigskip
\textsuperscript{1} Osnabrück University, Barbarastr. 12, 49076 Osnabrück, Germany
\\
\textsuperscript{2} Helmholtz-Centre for Environmental Research GmbH---UFZ, Permoserstr. 15, 04318 Leipzig, Germany
\\
\bigskip
* florian.schunck@uni-osnabrueck.de

\end{flushleft}

\section*{Abstract}
\begin{wrapfigure}[12]{r}{75mm}
    \includegraphics[width=75mm]{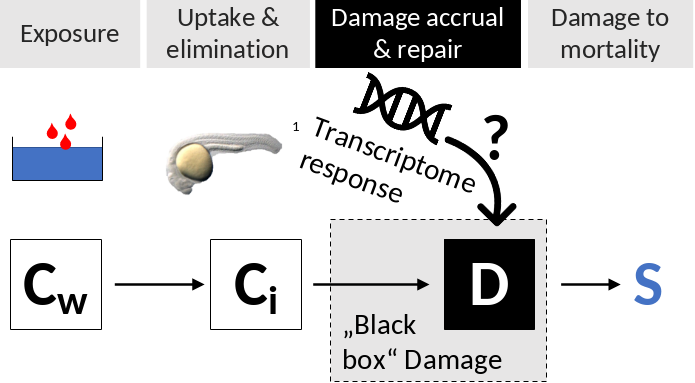}
\end{wrapfigure}
The immense production of the chemical industry requires an improved predictive risk assessment that can handle constantly evolving challenges while reducing the dependency of risk assessment on animal testing. 
Integrating 'omics data into mechanistic models offers a promising solution by linking cellular processes triggered after chemical exposure with observed effects in the organism. 
With the emerging availability of time-resolved RNA data, the goal of integrating gene expression data into mechanistic models can be approached. 
We propose a biologically anchored TKTD model, which describes key processes that link the gene expression level of the stress regulator \textit{nrf2} to detoxification and lethality by associating toxicodynamic damage with \textit{nrf2} expression.
Fitting such a model to complex datasets consisting of multiple endpoints required the combination of methods from molecular biology, mechanistic dynamic systems modeling and Bayesian inference. 
In this study we successfully integrate time-resolved gene expression data into TKTD models, and thus provide a method for assessing the influence of molecular markers on survival. 
This novel method was used to test whether, \textit{nrf2}, can be applied to predict lethality in zebrafish embryos. 
With the presented approach we outline a method to successively approach the goal of a predictive risk assessment based on molecular data.

\section{Introduction}\label{introduction}

The immense production of the chemical industry \cite{UNEP.2019} and the resulting release of novel substances into the environment \cite{Persson.2022} require an improved predictive risk assessment that can handle constantly renewing challenges. 
Tackling this problem experimentally has blind spots with respect to potentially vulnerable species (e.g.~pollinator decline \cite{Wood.2017,Beketov.2007,Rundlof.2015}, sublethal effects of chemicals and mixtures). 
The sheer combinatorial complexity of the problem precludes testing as a strategy. 
\textit{In-silico} approaches can be one way forward to achieve a prospective risk assessment and at the same time reduce an immense need for animal testing, if all of the above issues should be addressed. 
While data-driven approaches like QSARs and deep-learning display their power, they are constrained by the data they are calibrated to, hence extrapolation is limited \cite{Pearl.2019a,Pearl.2019}.
Mechanistic models encode causal relationships through processes over time, and are in this way capable of answering higher order questions such as ``What if?'', ``Why?'' or ``How would it look like under changed conditions?'' \cite{Pearl.2019}.
Such models can be designed when bio-physical processes are deciphered (advances in molecular biology), causal-relationships are defined through temporal sequences (mathematical abstraction), and data become available to learn the dominant processes that drive toxicity in humans or environmental organisms.

The growing availability of 'omics data drives the abstraction of bio-physical insights into the processes that govern molecular responses to changing environments \cite{Keenan.2018}.
The integration of 'omics data into mechanistic models therefore offers a promising solution for advancing risk assessment for chemicals and chemical mixtures, because in theory it can connect the cellular processes induced after toxicant exposure with observed effects in the organism \cite{Perkins.2019,Murphy.2018}.
Developing such approaches envisions the prediction of toxicant effects for untested species--substance combinations and mixtures as a very desirable long-term goal for a predictive environmental risk assessment.

It was recently shown that gene expression data from single time-point measurements \cite{Whitehead.2012} can be integrated into mechanistic models \cite{Stevenson.2023}. 
The next critical step is the integration of temporally-resolved 'omics data to also describe the dynamics of intermediate processes. 
To simultaneously model the dynamics of multiple process steps, models have to account for the relevant biological processes, thereby enhancing model accuracy and understanding of the intermediate processes that lead to the observed effects.
Compared to the efforts and costs of animal studies, that rarely provide time-resolved data, 'omics-assisted mechanistic models can offer a valuable and intriguing perspective for in-vitro bioassays. 

To advance prospective risk assessment, thus, the challenge arises to develop general models that are firmly grounded in biology and can integrate the temporal dynamics of multiple stages in the toxicant response, including molecular responses.

Toxicokinetic-toxicodynamic (TKTD) models seem ideally suited to facilitate the proposed integration of 'omics data into mechanistic models. 
They consider the uptake and elimination kinetics of chemicals (toxicokinetics, TK) and translate internal concentrations or other dose metrics to dynamical toxic effects (toxicodynamics, TD). 
TKTD models are frequently applied to model toxic effects over time \cite{Ashauer.2017}, and interactions between chemicals \cite{Cedergreen.2017,Singer.2023,Ashauer.2016,Dalhoff.2020}.
Particularly the general unified threshold model for survival (GUTS) \cite{Jager.2011} is a commonly used TKTD framework to model time-resolved survival data and even effects of chemical mixtures over time \cite{Bart.2022}.
Most importantly, GUTS models include a \emph{damage} state, which responds to internal toxicant concentrations and represents an impact state inside an organism from which observable effects follow.
This damage-state is abstract, but as well might be the state variable that corresponds most with 'omics data. Further investigation of the potential correspondence between the GUTS damage state and 'omics data is very challenging because of the limited availability of datasets that include both toxicokinetic, molecular and apical endpoints over time. In this study, we address this challenge and suggest a model structure that integrates gene expression data into TKTD models to approximate the damage state. 

A central pathway involved in the translation of environmental concentrations to observable effects is the integrated stress response (ISR)\cite{Gasch.2000}. It is an intracellular signalling network that helps cells and organisms to maintain health in a variable environment. It modulates cellular processes, among them mRNA translation and metabolism to enable cells to repair damage \cite{Costa-Mattioli.2020}, or if damage repair is unsuccessful, triggers apoptosis to remove damaged cells. 
In the cellular stress response to chemical exposure, Nrf2 has been identified as a master regulator of the detoxification process and its signalling pathway has been extensively described \cite{Zhao.2006,Raghunath.2018}.
Under basal conditions, \textit{nrf2} transcription and synthesis rates to Nrf2 proteins are kept in balance by KEAP1 proteins and ubiquitination targeted degradation \cite{Khalil.2015,Badenetti.2023} with a half-life of approximately 10--20 minutes \cite{Khalil.2015,Kobayashi.2004}. 
Nrf2 activation is tightly linked to the AhR pathway, that is also known to be one of the major chemical-induced metabolic pathways, and to the KEAP1 pathway, known to play a key role in oxidative stress response in organisms \cite{Li.2019}. 
Upon activation, Nrf2 translocates into the nucleus and activates transcription of genes that remediate stress via interaction with antioxidant response elements (ARE)s. 
Nrf2 activation can, therefore, be understood as a proxy indicative of stress induced by chemical exposure and related toxicity. 
In a recently published study, the transcriptome of zebrafish embryos (ZFE) was measured at multiple points in time, after exposure to toxicants \cite{Schuttler.2019}. 
The regulation frequency of various gene-clusters was temporally related to the internal concentration profiles in ZFE, indicating the value of gene-expression data in modeling the response to toxicants. 
Typical pulse-like expression profiles \cite{Yosef.2011,Bar-Joseph.2012} were observed in a gene cluster containing the \textit{nfe2L2b} gene (from here referred to as \textit{nrf2}), which expresses the Nrf2 protein in zebrafish, indicating that active reduction of damage needs to be considered in the modeling when integrating molecular responses such as \textit{nrf2} expression into TKTD models.

We hypothesize that toxicodynamic damage can be approximated by gene-expression data and thus serve as an interface for integrating 'omics data into mechanistic models.
To investigate this hypothesis, we use the time-resolved \textit{nrf2} expression signal from the published dataset of \cite{Schuttler.2019} complemented with unpublished time-series data of internal toxicant concentrations and survival and assess the potential of integrating it into TKTD models.

We further hypothesize that the expression of stress regulator gene \textit{nrf2} can be used to model lethality in zebrafish embryos independent from specific toxicant characteristics such as the mode of action. 
For testing this hypothesis,  we employ an approach of parameter sharing \cite{Singer.2023}, meaning that a combined model is fitted to data from testing  multiple substances with substance specific parameters for uptake, and general parameters for gene-expression and protein dynamics. Typically, TKTD models are parameterized on single-substance single-species datasets, but we propose that by integrating 'omics data as a damage proxy, this paradigm can be overcome for a subset of the parameters.

This study provides a biologically-anchored TKTD model, which links the gene expression level of \textit{nrf2} to detoxification and lethality, and in that way replaces the damage state in the standard GUTS models.
A complex dataset consisting of multiple fragmented endpoints was fitted with a stochastic variational inference algorithm, and the parameter uncertainties were thoroughly assessed.

\section{Methods}\label{methods}

\subsection{Description of the GUTS-RNA-pulse model}\label{description-of-the-reversible-damage-model}

\begin{figure}[ht!]
    \centering
    \includegraphics[width=\textwidth]{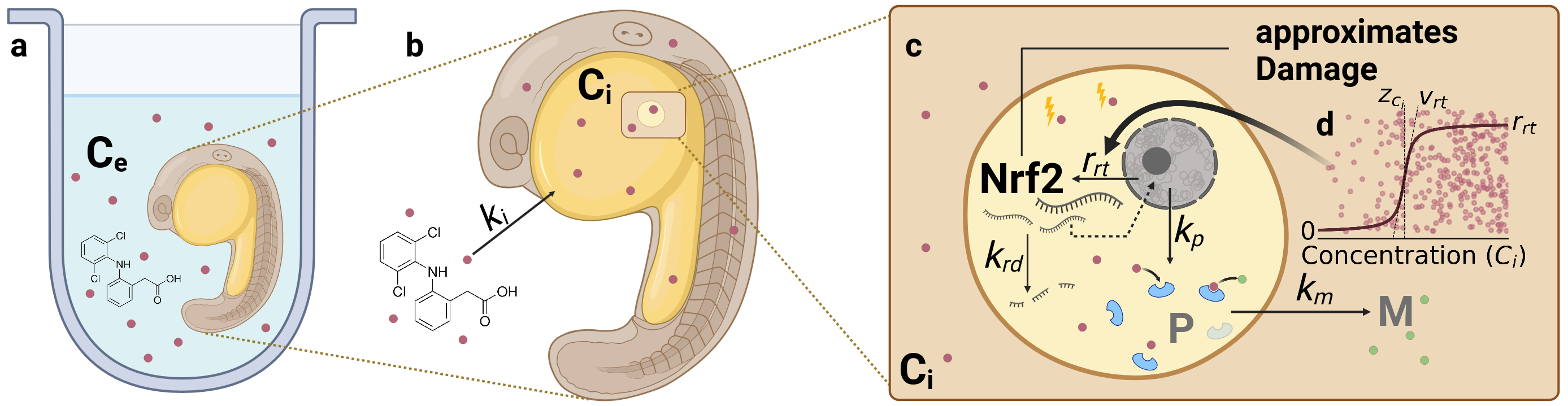} 
    \caption{
        Graphical description of the GUTS-RNA-pulse model, where the \textit{nrf2} concentration in the whole organism is used as a proxy for toxicodynamic damage. 
        a) Zebrafish embryo exposed to a chemical from 24—120 hours post fertilization (hpf). 
        b) Uptake of diclofenac (exemplarily) into the organism at rate constant $k_i$ \cref{eq:ci_dt}. 
        c) Zoom into the hypothesized expression metabolization process: \textit{nrf2} ($R$) is expressed at a constant rate $k_{rt}$ with a responsiveness $v_{rt}$ when the internal substance concentration $C_i$ exceeds a substance specific threshold $z_{ci}$ (shown in d) and decays at a concentration dependent rate constant $k_{rd}$ \cref{eq:nrf_dt}. 
        Unobserved metabolizing protein dynamics depend on the \textit{nrf2} concentration and are described with a dominant rate constant $k_p$ \cref{eq:protein_dt}. 
        Metabolization of the chemical depends on the protein concentration $P$, $C_i$ and the metabolization rate constant $k_m$ \cref{eq:ci_dt}.
        The non-monotonically increasing level of \textit{nrf2} is used to approximate toxicodynamic damage and is linked to ZFE survival via the stochastic death model by \cref{eq:survival_hazard,eq:survival_fct}. Created with \href{https://biorender.com}{BioRender.com}.
    }
    \label{fig:model_description}
\end{figure}
    
In this study, several toxicokinetic-toxicodynamic (TKTD) models are compared. 
Here, we focus on describing the integration of RNA expression into a GUTS TKTD model (named GUTS-RNA-pulse hereafter).
In the model, a constant \textit{nrf2} expression rate follows a concentration dependent activation \cref{fig:model_description}.
\textit{Nrf2} is assumed to be indirectly responsible for the metabolization of the chemical in the organism and is also linked to survival via a threshold model.

\paragraph*{\textit{nrf2} activation}\label{nrf2-activation}

As initially described the chemical stress response with respect to \textit{nrf2} regulation is complex.
There are several \textit{nrf2} regulating signalling pathways, including Nrf2 dissociation from KEAP1 and stabilization of Nrf2 proteins in the cytosol, Nrf2 autoregulation, and AhR induced response \cite{Li.2019}.
In order to generalize, the model focuses on commonly observed gene-expression patterns, which are (1) short expression impulses and (2) sustained expression \cite{Yosef.2011, Bar-Joseph.2012}.
Pulsed \textit{nrf2} expression was also observed in the data used in this study \cite{Schuttler.2019}.
In order to describe pulsed and sustained RNA dynamics a threshold activation model is proposed:

\begin{equation} \label{eq:nrf_dt}
    \frac{dR}{dt} = r_{rt}~\text{activation}(C_i,~C_{i,max},~ z_{ci},~ v_{rt}) - k_{rd} ~ (R - R_0)
\end{equation}

The model describes the relative differential transcription of RNA, denoted $R$, as a zero-order kinetic \cite{Qiu.2022,Xu.2023a} process with a constant transcription rate $r_{rt}$, activated when internal concentrations exceed a threshold $z_{ci}$ \cref{fig:model_description}. 
The slope of the activation is controlled by the parameter $v_{rt}$. 
The activation is any sigmoid function between 0 and 1. 
Inside the activation function $C_i$ is scaled with the maximum observed internal concentration over all experiments $C_{i,~max}$, in order to increase numerical stability of the function and to harmonize the scales of the $v_{rt}$ parameter \cref{si:eq:activation}. 
$R$ is degraded with first-order kinetics \cite{Chen.2008,Blake.2024} with the rate constant $k_{rd}$ when the initial RNA concentration $R_0$ is exceeded.
Here we assume $R_0 = 1$, which makes the modeled quantity identical to the measured fold change.
Note that this only applies when the baseline RNA expression is assumed constant, which it likely isn't in a developing organism.
For a deeper treatment of the relationship between differential RNA expression and measured fold change values, refer to \cref{si:sec:fold_change}.
The described differential equation model results in sustained gene-expression, when $C_i$ never falls below the threshold and it results in pulsed expression when the internal $C_i$ only transiently exceeds the internal concentration threshold.

\paragraph*{Uptake and metabolization} \label{uptake-and-metabolization}

\textit{nrf2} activation is linked to antioxidant response element (ARE) translation, especially, when activated via the AhR pathway, Nrf2 activates metabolization proteins that are involved in degrading the chemicals that activated the response.
Since also here, the data of the temporal dynamics of the stress response are limited, we assume that the metabolization process is described by the concentrations of internal concentration $C_i$, (metabolizing) protein $P$ concentration and a metabolization rate constant $k_m$, leading to an overall equation for the internal concentration 

\begin{equation} \label{eq:ci_dt}
    \frac{dC_i}{dt} = k_i~C_e - k_m~C_i~P
\end{equation}

The latter term can be understood as a Michaelis-Menten enzyme kinetic for relatively low substrate concentrations $C_i$.
All processes involved in the active detoxification rate are aggregated into a single quantity $P$, which changes depending on the \textit{nrf2} concentration and the metabolizing protein concentration with a dominant rate constant $k_p$. Passive detoxification, independent of the $P$ concentration, is not considered in this model.
This leads to the term 

\begin{equation} \label{eq:protein_dt}
    \frac{dP}{dt} = k_p~ ((R - R_0) - P)
\end{equation}

This equation is included to account for metabolization reactions that persist after transient gene-regulation pulses, based on the higher stability of proteins with half-lives between 20--46 hours \cite{Harper.2016} over \textit{nrf2} transcripts with approximated half-lives of 20 minutes \cite{Kobayashi.2004}.

\paragraph*{Survival} 

The survival probability $S$ is modeled according to the stochastic death assumption of the GUTS framework \cite{Jager.2011,Jager.2018a}, where the hazard is approximated by \textit{nrf2} fold-change. 

\begin{equation} \label{eq:survival_hazard}
    h(t) = k_k~ max(0, R(t) - z) +  h_b
\end{equation}

\begin{equation} \label{eq:survival_fct}
    S(t) = e^{-\int_0^t h(t) dt}
\end{equation}

\begin{table}[htb]
    \centering
    \footnotesize
    \begin{threeparttable}
        \caption{TKTD Parameters used in the GUTS-RNA-pulse model. The column ``Assumed substance independence'' indicates whether a parameter is supposed to be shared for multiple  substances.}
        \label{tab:parameters}
        \begin{tabularx}{\textwidth}{s X s s}
        \toprule
        Parameter              & Definition & Unit & Assumed substance independence \\
        \midrule
        ${k}_{i}$              & Uptake rate constant of the chemical into the internal compartment of the ZFE & $h^{-1}$ & no \\
        ${k}_{m}$              & Metabolization rate constant from the internal compartment of the ZFE& $\frac{L}{\mu mol~h}$ & no \\
        ${z}_{\text{ci}}$      & Scaled internal concentration threshold for the activation of \textit{nrf2} expression & $\frac{\mu mol~L^{-1}}{\mu mol~L^{-1}}$ & no \\
        ${v}_{\text{rt}}$      & Scaled responsiveness of the \textit{nrf2} activation (slope of the activation function) & $\frac{\mu mol~L^{-1}}{\mu mol~L^{-1}}$ & yes/no \tnote{\textit{a}} \\
        ${r}_{\text{rt}}$      & Constant \textit{nrf2} expression rate after activation \tnote{\textit{b}} & fc \tnote{\textit{c}} & yes \\
        ${k}_{\text{rd}}$      & Nrf2 decay rate constant & $h^{-1}$ & yes \\
        ${k}_{p}$              & Dominant rate constant of synthesis and decay of metabolizing proteins & $h^{-1}$ & yes \\
        ${z}$                  & Effect \textit{nrf2}-threshold of the hazard function \tnote{\textit{b}} & fc \tnote{\textit{c}} & yes \\
        ${k}_{k}$              & killing rate constant for \textit{nrf2} \tnote{\textit{b}} & $fc^{-1}~h^{-1}$ \tnote{\textit{c}} & yes\\
        ${h}_{b}$              & background hazard rate constant & $h^{-1}$ & yes \\
        $\sigma_{\text{cint}}$ & Log-normal error of the internal concentration & & yes \\
        $\sigma_{nrf2}$ & Log-normal error of the \textit{nrf2} expression \tnote{\textit{b}} & & yes \\
        \bottomrule
        \end{tabularx}    
        \begin{tablenotes}
            \item[\textit{a}] In an unscaled version of the activation function, $v_{rt}$ is not considered substance independent, due to an inverse relationship between $v_{rt}$ and $C_{i,max}$
            \item[\textit{b}] relative to the \textit{nrf2} concentration in untreated ZFE (fold-change)
            \item[\textit{c}] fold change: $\frac{\mu mol~nrf2\text{-treatment}~L^{-1}}{\mu mol~nrf2\text{-control}~L^{-1}}$ 
        \end{tablenotes}
    \end{threeparttable}
\end{table}

\paragraph*{Error models}

As error models we use log-normal distributions for \textit{nrf2} and internal concentration measurements to account for the fact that these values are constrained to the positive scale.
For survival data, a conditional binomial model was used, which is equivalent to the multinomial model for survival, which is the suggested likelihood function for estimating parameters for survival of small sampling groups with a repeated observations over time \cite{Ashauer.2016,Jager.2018a}. 

\paragraph*{Standard GUTS models}

The GUTS-RNA-pulse model was compared to additional guts model variants: GUTS-reduced (fitted only to survival data, \ref{si:mod:guts-reduced}), GUTS-scaled-damage (fitted to survival data and internal concentrations \ref{si:mod:guts-scaled-damage}), full GUTS (fitted to survival, internal concentration and \textit{nrf2} fold-change data, \ref{si:mod:guts-rna}, referred to henceforth as GUTS-RNA). These models have been described in detail \cite{Jager.2011,Jager.2018a} and will consequently not be further detailed in this study.

\subsection{Data description}\label{data-description}

\subsubsection{\textit{nrf2} data}

This study utilizes a published dataset (\url{https://academic.oup.com/gigascience/article/8/6/giz057/5505355}) of gene-expression time series of ZFE exposed to diuron, diclofenac and naproxen from 24 hours post fertilization (hpf) to 120 hpf \cite{Schuttler.2019}. The details on the underlying methods are available in the publication's supporting information. 

\subsubsection{$C_{ext}$, $C_{int}$ and survival data}\label{experimental-design-of-the-reversible-damage-project}

To achieve the goals of this work, the \textit{nrf2}-dataset has been complemented by time-resolved external concentration measurements ($C_{ext}$), internal concentration measurements ($C_{int}$) and apical effect observations. The data originate from a series of laboratory experiments conducted at the UFZ in Leipzig and are described in the following.

\textit{Test substances}
Active substances in the exposure experiments were diclofenac sodium salt (CAS: 15307-79-6, purity: n/a, batch: BCBP9916V, supplier: Sigma), diuron (CAS: 330-54-1, purity: 99.6\%, batch: SZBB265XV) and naproxen sodium salt (CAS: 26159-34-2, purity: 98-102\%, batch: MKBV4690V, supplier: Sigma). Exposure solutions were prepared freshly (\textless{} 24 h) for each experiment by dissolving pre-weighed amounts of diclofenac or naproxen in ISO-H2O (ISO 7346-3: 79.99 mM CaCl2*2H2O, 20.00 mM MgSO4*7H2O, 30.83 mM NaHCO3, 3.09 mM KCl; pH 7.4, oxygenized).
In case of diuron, exposure solutions were prepared from stocks with the substance dissolved in methanol (CAS: 67-56-1; purity: 100\%, batch: n/a, supplier: J.T. Baker). Final~solvent concentration in exposures and the respective controls was 0.1\%.
Test concentrations \cref{tab:si-experiments} were prepared in serial dilutions shortly before exposure initiation. 

\textit{Zebrafish handling and exposure}
For all experiments, eggs from zebrafish of OBI/WIK UFZ strain that have been reared under constant conditions throughout all included experiments (carbon-filtered tap water, 26°C, continuous aeration,  14:10 h light:dark cycle).
Within 30 min after spawning, eggs were collected, only~fertilised, and undamaged eggs in 4-32 cell stage pre-sorted. To ensure comparability to the gene-expression data set \cite{Schuttler.2019}, ZFE were incubated until exposure at 24 hpf. The following day, coagulated, damaged or developmentally delayed ZFE were discarded. Healthy ZFE were exposed by transferring three ZFE~with 50 µL ISO-H2O into a 7.5 mL glass vial prefilled with 6 mL exposure solution or control medium.
Six replicates were used in negative controls and for each substance concentration ZFE were exposed~in triplicates, with each replicate containing three organisms.
Pre-sorted as well as exposed ZFE were incubated in a climate chamber (Vötsch 1514, Vötsch Industrietechnik GmbH) at 26°C with a 12:12 h light:dark cycle on a shaker (Edmund Bühler SM-30 Control) with 75 rpm. Apical effects were observed under a stereo light microscope at 24, 48, 72 and 96 hpe.
Based on the results of acute toxicity tests, ZFE were exposed in additional sets of experiments at a concentration around the LC$_{25}$ derived for each test substance to determine internal concentrations ($C_{int}$).
Contrary to acute exposures, each replicate consisting of 10 ZFE was either exposed in 20 mL glass vials containing 18 mL exposure solution or control medium or 7.5 mL glass vials containing 6 mL, depending on the experiment (for details see \cref{tab:si-experiments}).

\textit{Sampling, Preparation, extraction and measurement of internal and external concentrations}
After 1.5, 3, 6, 8, 10, 12, 24, 36, 48, 60, 72, 84 and 96 hpe the replicates were pooled and samples for analysis of exposure concentrations taken. Due to the large number of experiments included in this dataset not all treatments included the same number of sampling times.
Dead or manually damaged organisms were discarded and from the remaining, 20 ZFE were randomly selected and transferred into a MP-tube. In case of internal concentration samples, ZFE were pipetted dry, rinsed once with 1 mL ISO-H2O and pipetted dry again.
All samples were immediately frozen in liquid N2 and stored at -20°C ($C_{int}$)) or -80°C (gene expression) until further processing. 
External concentration and homogenised internal concentration samples were measured with a liquid chromatography-high resolution mass spectrometry (LC-HRMS) system; for details see \cref{si:lcms-method}.

\subsection{Parameter estimation}\label{parameter-estimation}

One of the challenges for developing integrated TKTD models is the structure of the experimental data.
Although the work aims to make assessments of the biological processes within \textbf{one} organism, the experimental data are fragmented across many organisms, due to experimental necessity.
Thus, fitting the model on individual replicates is simply not possible, because replicates often include only one, sometimes two endpoints.
In this work, this dataset consisted of 202 treatments, 1426 observations distributed over 23 time points and 3 endpoints \cref{tab:si-experiments}.
The number of \emph{missing information} is several times larger than the number of observations \(202 \times 23 \times 3 - 1426 = 12512\).
To overcome this challenge, datasets from numerous biological experiments need to be combined together into a single large dataset, which is used to estimate the parameters of the described models.
Bayesian parameter inference approaches accommodate all these necessities and, further, report parameter distributions, reflecting the uncertainty in the true parameter and also provide estimates of the expected variation in experimental observations.
In this work, \textit{numpyro} \cite{Bingham.2018,Phan.2019,Bradbury.2018} was used as a probabilistic programming language (PPL) to efficiently ($\approx 40$~ms for solving the ODEs of all 202 experiments) estimate the distributions of the model parameters with Bayesian methods.
The state-of-the-art gradient-based Markov-chain Monte-Carlo (MCMC) sampler, NUTS \cite{Hoffman.2011}, was used to infer the parameters for the GUTS-reduced model and stochastic variational inference (SVI) \cite{Blei.2017} was used for all other models which were computationally more demanding. The exact details are described in \cref{si:sec:bayesian-parameter-inference}. 
Model development and parameter estimation were carried out with the modeling platform \texttt{pymob} (\url{https://github.com/flo-schu/pymob}), which allows seamless switching between parameter optimization/estimation algorithms. To assess parameter uncertainty and identify  identifiability issues, 100 estimations were started with initial parameters drawn from a uniform interval from -1 to 1, which were subsequently transformed to the scales priors of the parameter distributions. The exact algorithm is described in \Cref{si:sec:parameter_analysis_algorithm}.

\section{Results}\label{results}

\subsection{Using time resolved \textit{nrf2} data in combination with a reversible damage dynamic in TKTD models is possible and can correctly describe the dynamics of survival.}

In the first step of this work, 4 different models were fitted on the available data on a per substance basis (GUTS-reduced, GUTS-scaled-damage, GUTS-RNA and GUTS-RNA-pulse).
The GUTS-scaled-damage model serves as a baseline and was solely fitted on internal concentration and survival data.

\begin{figure}[htb]
    \includegraphics[width=\textwidth]{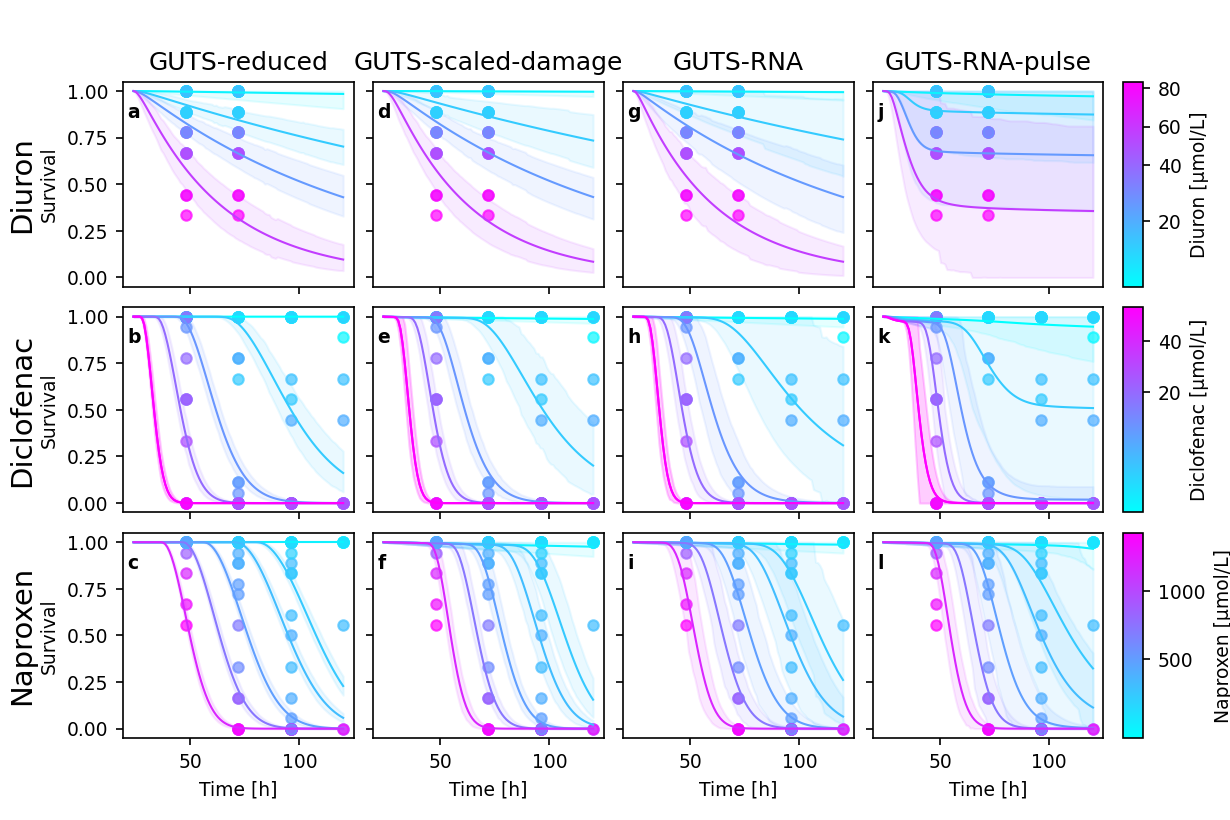} 
    \caption{Survival under exposure to different chemicals (rows) for the different GUTS model flavours (columns) over time, where t=0 is the moment of fertilization of the egg.
    a--c: GUTS-reduced model fits on  survival $S$ data only. 
    d--f: GUTS-scaled-damage mode fits on internal concentrations $C_i$ and $S$.
    g--h: GUTS-RNA (full GUTS) model fits on $C_i$, $D$ with \textit{nrf2} expression as a damage proxy and $S$.
    j--l: GUTS-RNA-pulse model fits on $C_i$, $D$ with \textit{nrf2} expression as damage proxy and $S$.
    The rows display different substances (diclofenac, diuron, naproxen).}
    \label{fig:survival_comparison}
\end{figure}

Survival after exposure to diclofenac and naproxen can be successfully modeled with the GUTS-scaled-damage (Fig.~\ref{fig:survival_comparison} e, f) and the GUTS-reduced (Fig.~\ref{fig:survival_comparison} b, c) model. 
However, both models are unable to fit constant survival over time at continuous exposure to diuron (Fig.~\ref{fig:survival_comparison} a, d). \\
The GUTS-RNA model uses the fold-change values of \textit{nrf2} gene expression as a proxy for the damage state.
This integration does not affect the matching of modelled and observed survival for the different substances (Fig.~\ref{fig:survival_comparison} g--h), but it reduces the variability in the killing rate $k_k$ and threshold $z$ parameters of the stochastic death model across substances \cref{si:tab:parameters-guts-rna-pulse}.
In this sense, the integration of \textit{nrf2} expression data into a TKTD model is already successful.
In general, the large variability in observed data, introduced by processing over 200 treatments in one model, is reflected in the large uncertainty intervals of the model fits.
Contrary to intuition, the uncertainty further increased with an increasing number of qualitatively different observations (survival, internal concentrations, gene-expression) in the model.
This suggests that the confidence in parameter estimates expressed by the more reduced models, may be misleading.

\subsection{Modeling damage as a reversible process allows describing constant survival at continuous exposure}\label{modeling-damage-as-a-reversible-process-can-model-constant-survival-at-continuous-exposure}

To improve modelling constant survival over time under constant exposure, the GUTS-RNA-pulse model was developed (see model description) and fitted to the data.
The model implements an RNA-protein dynamic \cref{eq:nrf_dt,eq:protein_dt} that approximates key biological processes in the stress response.
Only the GUTS-RNA-pulse can accurately reproduce the survival dynamic in all three investigated substances (Fig.~\ref{fig:survival_comparison} j--l). This can be explained by the ability of the GUTS-RNA-pulse approach to model transient damage pulses, meaning reversible damage at constant exposure.
The improved matching of the model comes at costs of an increase in the number of parameters and an increased uncertainty.
While the full GUTS-RNA model needs 7 parameters (excluding the error parameters), the GUTS-RNA-pulse model requires 10 parameters.
However, those parameters are connected to explicit biological meanings (\textbf{Table 1}).

\begin{figure}[htb]
    \centering
    \includegraphics[width=\textwidth]{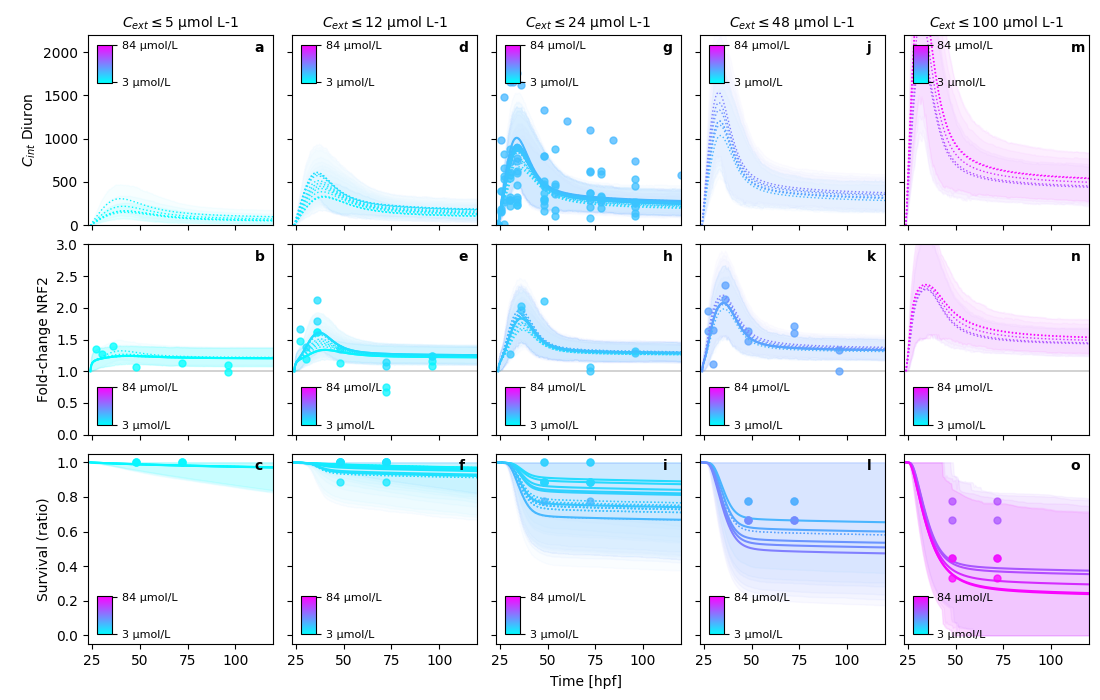} 
    \caption{Exemplary model fits and posterior predictions of the GUTS-RNA-pulse model for diuron. 
    To improve the readability of the figure, columns distribute the modeled experimental data into concentration classes.
    The solid lines are the mean posterior predictions of the endpoints over time and the dotted lines are those predictions where no data were available.
    The shaded areas indicate the posterior density intervals containing 95\% probability of the posterior predictions.
    Note that the uncertainty in the observations is not included in these figures.}
    \label{fig:guts_rna_pulse_diuron}
\end{figure}

The value of the added reversible \textit{nrf2} dynamics in terms of an RNA-concentration dependent, activated expression and metabolization is showing particularly well in the case of the diuron exposure \cref{fig:guts_rna_pulse_diuron}.
Here, the dynamics of \textit{nrf2} can be described very well over a wide range of concentrations.
In addition, extrapolation of \textit{nrf2} predictions to concentration ranges that were not measured looks promising, as the model limits \textit{nrf2} expression even at high concentrations to reasonable ranges.
This behaviour of the GUTS-RNA-pulse model is similar for all investigated compounds (Figs.~\ref{si:fig:model_fits_guts_rna_pulse_specific_diclofenac},~\ref{si:fig:model_fits_guts_rna_pulse_specific_naproxen}).
In contrast, at high concentrations, the GUTS-RNA model predicts very high \textit{nrf2} expression (Figs.~\ref{si:fig:model_fits_guts_rna_diuron}--\ref{si:fig:model_fits_guts_rna_naproxen}).

In addition to the improved description of the \textit{nrf2} dynamics, also the internal concentration dynamics of diuron (Figs.~\ref{fig:guts_rna_pulse_diuron} a, d, g, j, m) and naproxen (Fig. \ref{si:fig:model_fits_guts_rna_pulse_specific_naproxen} a, d, g, j, m, p) are more reasonably described by the RNA-pulse model compared to other GUTS variants (Figs.~\ref{si:fig:model_fits_guts_rna_diuron}--\ref{si:fig:model_fits_guts_reduced_naproxen}).
This supports the assumption that active metabolization of chemicals under constant exposure has likely been a relevant process in these experiments Figure~\ref{fig:survival_comparison}.

Evidently, also the GUTS-RNA-pulse model is not complete.
Despite the improved description of survival and internal concentration, the metabolization kinetics of diuron are overestimated (Figs.~\ref{fig:guts_rna_pulse_diuron} a, d, g, j, m).
This can be explained by the very abrupt change in the accumulation rate of diuron in the ZFE.
Coupled with the described signalling pathway $C_i^{\uparrow} \rightarrow \textit{nrf2} \rightarrow P \rightarrow C_i^{\downarrow}$, only a very fast metabolization rate can compensate for the induced time lag between arrival of the compound in the internal compartment and the onset of metabolization.
This leads to an metabolization overshoot, which explains the modeled kinetics. \\
Also for diclofenac, observed \textit{nrf2} and $C_i$ dynamics are very complex \cref{si:fig:model_fits_guts_rna_pulse_specific_diclofenac}.
If individual concentration trajectories are inspected, multiple enrichment phases are visible: Fast initial uptake is followed by slowed uptake, and followed by reduction of the internal concentration.
Such dynamics are not yet possible to be modelled with a \emph{simple} RNA pulse model. 
Obviously the true dynamics of the signalling pathways of the stress response are much more complicated, and has been modelled with more than 30 ODE terms \cite{Hiemstra.2022}.

Despite these limitations, the proposed GUTS-RNA-pulse model has several advantages. 
First of all, it provides evidence that it is indeed possible to integrate gene expression data into TKTD models; therefore we maintain hypothesis 1.
In addition, it provides a possible solution to the problem of modeling reversible damage and predicting constant survival at continuous exposure to toxic chemicals. 
Further, the RNA-protein dynamic \cref{eq:nrf_dt,eq:protein_dt} should be independent of the exposed substance and with them the coupled stochastic death model.

\subsection{Survival dynamics can not solely be predicted by an \textit{nrf2} damage proxy}

After having established the integration of time-resolved \textit{nrf2} data into a TKTD model, we go on to test the hypothesis that the level of \textit{nrf2} expression predicts lethality in zebrafish embryos in a general way, independent from the specific substance.
For this we take the GUTS-RNA-pulse model and allow sharing of 7 parameters among the 3 substances in the parameter estimation, leaving only the parameters controlling uptake $k_i$, metabolization $k_m$ and the gene-expression activation threshold $z_{ci}$ as substance specific parameters. Hereafter, the parameter-sharing model will be referred to as the substance-independent model.
The substance-independent model has $7 + 3 \times 3 = 16$ parameters (excluding error parameters), compared to the substance specific models (\cref{fig:guts_rna_pulse_diuron,si:fig:model_fits_guts_rna_pulse_independent_diclofenac,si:fig:model_fits_guts_rna_pulse_independent_naproxen}), which have together $3 \times 10 =30$ parameters.
If \textit{nrf2} was a valid predictor (precursor) for lethality, and the model dynamic was accurately described, the model with fewer parameters should describe the lethality equally well.

\begin{figure}[htb]
    \centering
    \includegraphics[width=\textwidth]{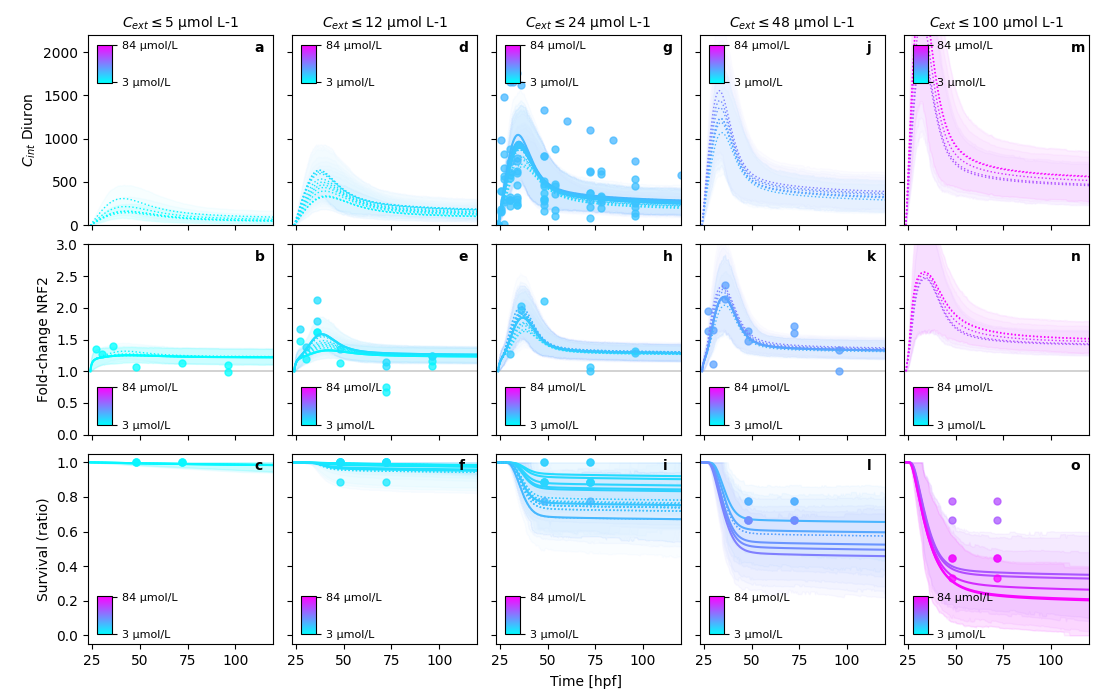}
    \caption{Exemplary model fits and posterior predictions of the parameter sharing (substance-independent) GUTS-RNA-pulse model for diuron.
    In order to improve the readability of the figure, columns distribute the modeled experimental data into concentration classes.
    The solid lines are the mean posterior predictions of the endpoints over time and the dotted lines are those predictions where no data were available.
    The shaded areas indicate the posterior density intervals containing 95\% probability of the posterior predictions.
    Note that the uncertainty in the observations is not included in these figures.}
    \label{fig:rna_pulse_independent_diuron}
\end{figure}

The substance-independent model indeed delivers accurate model fits for all observed endpoints after diuron exposure \cref{fig:rna_pulse_independent_diuron}.
Diclofenac fits of the substance-independent model \cref{si:fig:model_fits_guts_rna_pulse_independent_diclofenac} are also very similar to the fully substance specific model, although the predicted \textit{nrf2}-expression pattern is not in exact agreement with the data similar to the fully-substance specific model \cref{si:fig:model_fits_guts_rna_pulse_specific_diclofenac}.

Fits of naproxen data with the substance-independent model indicate that \textit{nrf2} gene expression data is not sufficient to predict lethality in zebrafish embryos \cref{si:fig:model_fits_guts_rna_pulse_independent_naproxen}.
The observed accurate fits for lethality can only be achieved by large deviations of the model fits from internal concentrations measurements and \textit{nrf2} expression measurements \cref{si:fig:model_fits_guts_rna_pulse_independent_naproxen}.
This deviation is underlined by the higher total Bayesian information criterion (BIC) value (for all three substances) of the substance-independent model (BIC = 6337) compared to the substance-specific model (BIC = 5940). 
Overall, the fits of the substance-independent model were surprisingly good, despite having 14 parameters less than the  substance-specific model. 
Nevertheless, our results show that our second hypothesis cannot be confirmed and \textit{nrf2} gene expression alone is not sufficient to predict lethality with the developed model.
\textit{nrf2} is known to play a key role in the induction of stress response including metabolization and detoxification. The observed data patterns underline this especially for diclofenac and diuron.
\subsection{Sharing parameters of the RNA-protein dynamic between different substances eliminated the problem of parameter identifiability.}

\begin{figure}
    \centering
    \includegraphics[width=.8\textwidth]{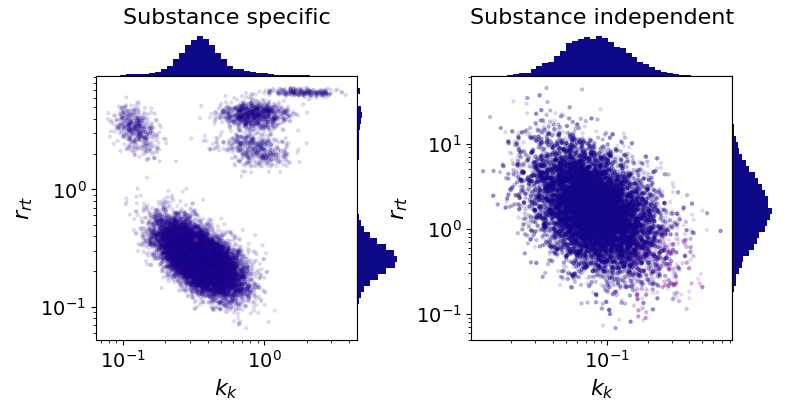}
    \caption{Joint posterior parameter distribution of $k_k$ and $r_{rt}$ for the diclofenac fit. 
    The model converges on multiple clusters of parameter estimates with very similar likelihood. 
    Left: Substance specific model. 
    Right: Substance independent model.}
    \label{fig:parameter_identifiability}
\end{figure}

Due to the model complexity, the GUTS-RNA-pulse model suffered from parameter identifiability issues.
This is exemplary shown in Figure~\ref{fig:parameter_identifiability}~a for the case of diclofenac. The parameter clusters have slightly different likelihoods, which originate from the deviation of parameter estimates from the prior probabilities. While these issues could be cured by simplifying the model, they could also be tackled by using more concentrated prior probabilities. The GUTS-RNA-pulse model allows this, because here most parameters have very specific biological meanings and could in theory be informed by empirical evidence from previous experimental work or expert knowledge.

Here, the problem was solved by making the RNA-protein dynamic substance independent (Fig.~\ref{fig:parameter_identifiability} b), which reduced the number of parameters from 30 to 16 and consequently the  uncertainty in parameter estimates.
The parameter inference approach applied in this work has great value, as it joins scattered datasets and extracts information from them, by combining effect measurements from start to finish.
It would not be possible to fit models separately on individual experiments, because the observed endpoints never jointly occur in a single experiment as all \emph{material} from the exposed organism were consumed for the separate analyses.

\section{Discussion}\label{discussion}

\subsection{Time resolved gene-expression data can be integrated into TKTD models as a proxy for toxicodynamic-damage}

The GUTS-RNA-pulse model described in this work delivered model fits of high quality with respect to the dynamics of the observed endpoints. We therefore maintain our hypothesis that time resolved gene expression data can be integrated into TKTD models.
While it is generally accepted that increasing the number of observations reduces uncertainty, in this work we observed the contrary for the successive increase in the number of endpoints added to the model. Each addition added uncertainty to the model and peaked when \textit{nrf2} gene expression data were included.
This observation showcases that adding qualitatively different information (as opposed to increasing the quantity) imposes constraints on the model, resulting in increasing the uncertainty in parameter estimates. This highlights that uncertainty estimates of reduced GUTS models, based solely on survival data, reflect only the model parameters' uncertainty and not the structural uncertainty.

\subsection{Hierarchical error modelling for better identifiability of model parameters}
\label{sec:discussion-error-modelling}

In the present work we used a complete pooling approach, i.e.~fitting the same model on all data, as opposed to using a hierarchical approach where different experiments may have varying parameter estimates.
While this complete pooling approach is easier to implement it has some drawbacks.
Since the data used in this work originated from multiple experiments, it contains complex nested error structures with 3 layers of errors:
\begin{enumerate}[topsep=0pt,itemsep=-1ex,partopsep=1ex,parsep=1ex]
    \item Experimental errors, e.g. biological batch differences between experiments, or variation in prepared stock solutions.
    \item Treatment errors, e.g. pipetting errors.
    \item Random (white noise) errors, e.g. measurement errors, or biological variation within the same batch.
\end{enumerate}
These errors levels are present in different depths of observations depending on the observed endpoint (repeated observations of the same individual in lethality measurements vs. independent measurements of \textit{nrf2} and $C_i$ data over time).

In the current approach, the same model will be forced on observations from different experiments and treatments; if the initial conditions (i.e. exposure) differ between experiments due to batch errors or experimental handling between experimentators, this will result in fitting a model that has to satisfy heterogeneous data with a single parameter per replicate or experiment. In the case of non-linear dynamics this can lead to biased estimations of model parameters \cite{Helleckes.2022}.
A better alternative is a partial pooling (hierarchical) approach where group level means are fitted which themselves serve as priors for the parameter estimates of individual experimental time series \cite{Helleckes.2022}.
Finding an accurate representation of the error structure and implementing it in a Bayesian parameter estimation scheme also allows the use of fixed error parameters. In the current approach, error parameters $\sigma$ \cref{tab:parameters} were fitted from the data and include noise from all error sources mentioned above. However, often knowledge exists about the variability of measurement noise.
Overall, a hierarchical approach allows a more exact attribution of errors to different levels of hierarchy and allows the use of fixed error parameters for e.g. measurement errors, which allows the model to fit closer to the intermediate endpoints ($C_i$, \textit{nrf2}) and yield more robust uncertainty estimates for the survival fits.

\subsection{\textit{nrf2}'s role in the stress response from a TKTD perspective}
\label{sec:discussion-nrf2-bio}

While the biology of the ISR is very complex and involves numerous genes, proteins and enzymes \cite{Gasch.2000,Costa-Mattioli.2020,Badenetti.2023,Hiemstra.2022}, the development of generalized models requires simplifications.
In the presented model, we focused on \textit{nrf2} and let it play a double role as a proxy for toxicodynamic damage and downstream trigger of biotransformation and detoxification. We tested whether \textit{nrf2} is a general predictor for lethality for any of the tested substances and concentrations. 
With the here described model, this hypothesis could not be confirmed.
This answer follows from the reduction of model fits for internal concentrations and \textit{nrf2} data when using substance independent parameters for the RNA-protein dynamic \cref{eq:nrf_dt,eq:protein_dt}.
Nevertheless, we found ample evidence that \textit{nrf2} is closely related to the process of activating metabolization and in that sense at least indirectly linked to lethality.

By modelling active metabolization, the phenomenon of reversible damage at constant exposure to toxicants was addressed.
Conventional TKTD models include no processes for reduction of damage that is not controlled by reductions in external concentration.
The process of reduction of internal concentration before the end of the exposure has been observed in ZFE after exposure to diazinon exposure at 0.4 ppm \cite{Keizer.1991}.
In the same study, exposure to 1/4 of the concentration did not lead to decreasing concentrations before the end of the exposure. 
In a different study, decreasing internal concentrations at constant exposure levels were observed for a variety of compounds such as benzocaine, phenacetin, metribuzin, phenytoin or valproic acid \cite{Brox.2016}; while the size increase of the developing organism could not account for the decrease in internal concentrations.
In another exposure study of $\alpha$-cypermethrin on \textit{Daphnia magna} \cite{Cedergreen.2017}, 
modelling active metabolization led to excellent model fits with constant survival predictions for diuron \cref{fig:guts_rna_pulse_diuron}, which indicates that a threshold model for active metabolization of a compound is indeed relevant. It was previously shown that temporal gene transcription changes of cytochrome P450 enzymes after exposure to benz[$\alpha$]anthracene follow pulse-like, fluctuating, or monotonically increasing dynamics that were additionally sensitive to the developmental age of ZFE \cite{Kuhnert.2017}.
Following this, temporally resolved concentrations of metabolites or RNA expression data for metabolic enzymes should be integrated into the proposed model to couple active metabolization with RNA expression by using advanced kinetic models such as, e.g. Michaelis-Menten kinetics. This could significantly improve the model, by considering self-reinforcing feedback loops from toxic metabolites which allows the modelling of fluctuating RNA expression patterns and multiphase-internal concentration and survival dynamics.

The substance-independent GUTS-RNA-pulse model estimates an RNA decay rate constant of $1.3/t$  (0.1--3.3).
This translates to a half-life distribution with a mode at 23 minutes and highly probable values between 0 and 1.5 hours (Fig.~\ref{si:fig:rna-half-life}).
This is in excellent agreement with experimentally established of RNA half-lives  of 10-20 minutes \cite{Kobayashi.2004,Khalil.2015}. Although protein kinetics were only calibrated indirectly, the estimated protein stability (half-life (mode) $\approx 46$~h) agreed with reported ranges of proteins between 20--46 h \cite{Harper.2016}, although the estimated half-life had wider tails towards longer half-lives \cref{si:fig:half-life-protein}. 
The agreement of fitted RNA half-life and protein half-life with reported literature data indicates that it is possible to use informative priors for RNA decay in future work.
The strong differences in molecular kinetics underline the importance that the temporal dimension has to be considered to understand the relationship between transcription and enzyme activity \cite{Glanemann.2003} and diverging measurements \cite{Velki.2017}.

Adding \textit{nrf2} expression as mechanistic information to the TKTD model can increase the interpretability of the killing rate parameter: Because \textit{nrf2} appears in the hazard function, the unit of $k_k$ in an unscaled model version would be $L~\mu mol~nrf2^{-1}~h^{-1}$.
Because Nrf2 is known to induce cell death (apoptosis) when the stress cannot successfully be reduced \cite{Costa-Mattioli.2020}, the killing rate parameter can be interpreted as the volume of dead cells per \textit{nrf2} fold-change increase above the threshold per unit time.
However, this interpretation is only possible when information of the \textit{nrf2} expression of the untreated organism is included, leading to the replacement of \textit{fold-change} with $\mu mol~L^{-1}$.

\section{Outlook}\label{outlook}

The presented model is far from being perfect and not close to the goal of reaching a predictive risk assessment without the need for animal testing. 
However, we are very confident to have shown a method that has immense potential to bring mechanistic understanding into the risk assessment.
We see two major advantages arising from this: 
Better extrapolation through incorporation of further biological processes and a reduced risk for overfitting, by increasing the uncertainty through inclusion of additional biological endpoints.
And, the ability to formulating experimentally falsifiable hypotheses and identification of clear molecular targets for testing the hypotheses (models) with the goal of successively improving model predictions.

Multiple concrete research directions are envisioned to proceed with maturing this approach.

\begin{enumerate}
    \item Deepening the model detail with respect to the biotransformation processes and enzyme kinetics to extend the model with respect to repair and rescue capacity.
    \item Improving the error modelling to account for experimental variation and improve uncertainty assessment.
    \item Integrate further molecular endpoints, such as oxidative stress markers \cite{Yang.2024}, or whole transcriptome expression data \cite{Schuttler.2019} into the model. However, such a process has high computational demands. 
    \item Make use of Bayesian parameter inference, by integrating prior knowledge into protein dynamics. It is known that proteins have a significantly larger half-life than RNA (20-46h) \cite{Harper.2016}. 
    \item Model sublethal effects, which are readily available for ZFE and can be investigated with the same approach with the potential to reduce animal testing. A respective database and software (INTOB) for systematic and FAIR ZFE phenotype data, including sublethal effects, will be available soon and might provide a valid prerequisite.  
    \item Integrate molecular information of the untreated organism into the model to move from a scaled model version (relative to the baseline organism) to a untreated organism to account for developmental processes within the lifetime of the zebrafish embryos.
    \item Validate the model against other substances in order to build confidence into the model or identify deficiencies.
    \item Extend the model to describe different types of typically observed RNA expression patterns such as fast-forward loops, cascades, autoregulation-loops, etc. \cite{Yosef.2011}. For instance, Nrf2 may induce KEAP1 expression which might limit uncontrolled \textit{nrf2} expression \cite{Li.2019}.
\end{enumerate}

\section*{Acknowledgements}
The authors thank Leonhard Bürger and Simon Hansul for constructive feedback on the model development, 
Anton Stratmann and Michael Osthege for important advice on the implementation of a Bayesian model,
Samuel Leweke for getting started with numpyro. Computations on the HPC of Osnabrück University were funded by the Deutsche Forschungsgemeinschaft
(DFG, German Research Foundation) - 456666331.


\printbibliography

\section*{Supporting information}\label{supporting-information}

\begin{itemize}
  \item The code used to generate the results including install and usage instructions are available on \url{https://github.com/flo-schu/tktd_nrf2_zfe/}
  \item A modeling notebook aligned with TRACE standards is available on \url{https://github.com/flo-schu/tktd_rna_pulse/blob/main/scripts/tktd_rna_3_6c_substance_specific.ipynb}
  \item The GUTS-RNA-pulse models are available on \url{https://github.com/flo-schu/tktd_rna_pulse}
  \item The standard GUTS models (reduced, scaled-damage, full GUTS) are available on \url{https://github.com/flo-schu/guts}
  \item The code for fitting parameters of the model is available on \url{https://github.com/flo-schu/pymob}
  \item The datasets are available on \url{https://doi.org/10.12751/g-node.l6jqgf} \url{https://gin.g-node.org/flo-schu/tktd_nrf2_zfe__data}
  \item Additional results for the GUTS-RNA-pulse models are available on \url{https://gin.g-node.org/flo-schu/tktd_rna_pulse__results}
  \item Additional results for the standard GUTS models (reduced, scaled-damage, full GUTS) models are available on \url{https://gin.g-node.org/flo-schu/guts__results}
\end{itemize}


\setcounter{subsection}{0}
\renewcommand{\thesubsection}{S\arabic{subsection}}
\setcounter{equation}{0}
\renewcommand{\theequation}{S\arabic{equation}}
\setcounter{figure}{0}
\renewcommand{\thefigure}{S\arabic{figure}}
\renewcommand{\thetable}{S\arabic{table}}

\subsection{Activation function for RNA expression}\label{si:sec:activation_fct}

\begin{equation}\label{si:eq:activation}
    activation(C_i,~C_{i,max},~z_{ci}, ~v_{rt}) = 0.5 + \frac{1}{\pi} ~ arctan(v_{rt} ~ (\frac{C_i}{C_{i,max}} - z_{ci}))
\end{equation}
which is numerically stable at high differences between $C_i$ and the threshold $z_{ci}$, in contrast to the conventionally used logistic function. For the activation calculation, $C_i$ was scaled by the maximum internal concentration $C_{i,~max}$ , in order to make threshold and slope parameters comparable across substances. In addition, this approach makes it also easier to use other activation functions like the logistic function.

\subsection{Interpreting fold-change quantities in the context of a dynamic model}\label{si:sec:fold_change}

Gene expression measurements in ZFE are usually based on calculating the differential expression (fold-change) of the treatment to the baseline and are typically reported as fold-change or log fold-change. 
The transformation of gene expression data (e.g.~$\Delta$, fold-change) has different pros and cons. The true effect is the effect that is easiest to model as it requires the least amount of assumptions, since it models only the effect, which should follow a relatively simple scheme \cite{Yosef.2011,Bar-Joseph.2012}. 

\begin{center}
    \includegraphics[width=0.5\textwidth]{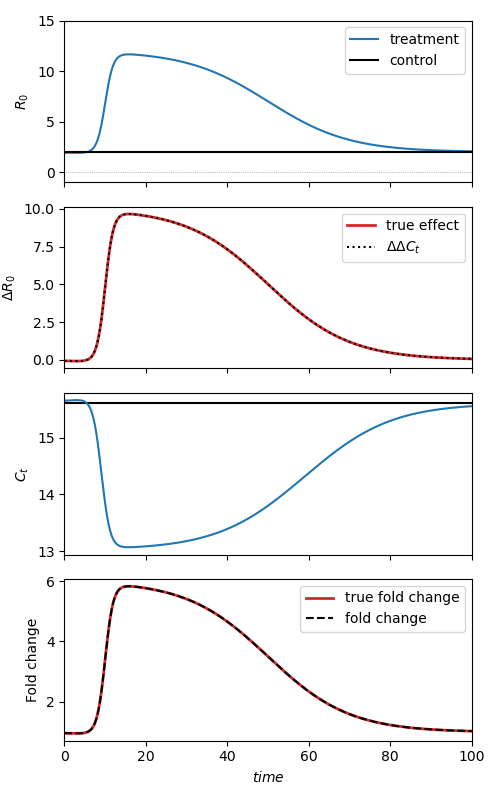}
    \captionof{figure}{Panel 1: Control (untreated) RNA expression rate and the treatment RNA expression rate as the combined signal of effect and control. Panel 2: true differential expression and measured differential expression. Panel 3: $C_t$ values over time of control and treatment. Panel 4: Modelled fold-change and measured fold change}
    \label{si:fig:fold_change}
\end{center}

The disadvantage is only proportional to the true effect if not scaled by measurement specific information such as the \textbf{threshold} value of a qPCR analysis at which the $C_t$ signal is obtained. The comparison of modeled \cref{si:eq:true-fold-change} and measured data \cref{si:eq:measured-fold-change} shows that the measured fold change is equal to the true (modeled) fold change \cref{si:fig:fold_change}.

\begin{equation} \label{si:eq:true-fold-change}
    \text{true fold change} = \frac{R_\text{treatment}} {R_\text{control}}
\end{equation}

\begin{equation} \label{si:eq:measured-fold-change}
    \text{measured fold change} = 2 ^ {-(C_t^\text{treatment} - C_t^\text{control})}
\end{equation}

The major disadvantage of using fold-change data is that a baseline has to be modeled implicitly if the true signal should be modeled recovered. Of course simple baseline models can be used such as a constant signal, increasing baseline (linear), step \ldots{}. In addition the baseline could be modeled as a stochastic function, such as a random walk where information from the dynamic of the baseline could come from the raw data of the controls.

If the baseline RNA expression is modeled explicitly, but data are only available as fold change, we recommend rescaling the RNA expression data according to \cref{si:eq:rna_fc}.

\begin{equation} \label{si:eq:rna_fc}
    R_{\text{fc}}(t) =  \frac{R(t) + R_{\text{control}}(t)}{R_{\text{control}}(t)}
\end{equation}

\subsection{LC/MS measurement method for internal and external concentrations} \label{si:lcms-method}
Diuron, diclofenac and naproxen were analysed by liquid chromatography-high resolution mass spectrometry (LC-HRMS) using a Ultimate 3000 LC system (Thermo) coupled to a Thermo LTQ Orbitrap XL We used a Kinetex Core-Shell C18 column (50 mm × 2.1 mm; 2.6 µm particle size; Phenomenex) and a gradient elution was carried out with a flow rate of 0.3 mL/min with water (A) and methanol (B) both containing 0.1\% formic acid. The initial content of 20\% B was linearly increased after 0.5 minutes to 100 \% B within 5.5 minutes. B was maintained at 100\% for 8 minutes followed by a re-equilibration for 5 minutes. The injection volume was 5 µL and the column was maintained at 40°C. A heated electrospray ionisation source was used in positive mode. Full scan spectra were acquired in centroid mode in a range of 80 to 600 m/z at a nominal resolving power of 30,000 referenced to m/z 400. A mass accuracy <7 ppm was assured over the whole mass range by external mass calibration using a calibration solution for the range from 138 to 1721 m/z.

\subsection{Experiments}

\begin{longtable}{lrrrlrrrr}
\caption{Included experiment in the present study. $C_{\text{ext}}$ refers to the nominal external concentration. $V_{\text{expo}}$ is the volume of the exposure solution in ml. $N_{\text{ZFE}}$ is the number of zebrafish embryos used in a single experimental replicate. $N_{\text{Trt}}$ is the number of treatments applied for the given exposure concentration.$N_{\text{Obs}}$ refers to the cumulative number of observations for all replicates and observation times.} \label{tab:si-experiments} \label{si:tab:experiments} \\
\toprule
Endpoint & ID$_{\text{Exp}}$ & Year & $N_{\text{ZFE}}$ & Substance & $C_{\text{ext}}$ & $V_{\text{expo}}$ & $N_{\text{Trt}}$ & $N_{\text{Obs}}$ \\
\midrule
\endfirsthead
\caption[]{Included experiment in the present study. $C_{\text{ext}}$ refers to the nominal external concentration. $V_{\text{expo}}$ is the volume of the exposure solution in ml. $N_{\text{ZFE}}$ is the number of zebrafish embryos used in a single experimental replicate. $N_{\text{Trt}}$ is the number of treatments applied for the given exposure concentration. $N_{\text{Obs}}$ refers to the cumulative number of observations for all replicates and observation times.} \\
\toprule
Endpoint & ID$_{\text{Exp}}$ & Year & $N_{\text{ZFE}}$ & Substance & $C_{\text{ext}}$ & $V_{\text{expo}}$ & $N_{\text{Trt}}$ & $N_{\text{Obs}}$ \\
\midrule
\endhead
\midrule
\multicolumn{9}{r}{Continued on next page} \\
\midrule
\endfoot
\bottomrule
\endlastfoot
$C_{\text{int}}$ & 21 & 2016 & 9 & diclofenac & 5.0 & 18 & 2 & 34 \\
$C_{\text{int}}$ & 21 & 2016 & 9 & diclofenac & 6.6 & 18 & 3 & 35 \\
$C_{\text{int}}$ & 22 & 2016 & 9 & diclofenac & 6.6 & 18 & 6 & 65 \\
$C_{\text{int}}$ & 22 & 2016 & 9 & diclofenac & 7.2 & 18 & 6 & 70 \\
$C_{\text{int}}$ & 25 & 2017 & 9 & diclofenac & 6.6 & 18 & 2 & 44 \\
$C_{\text{int}}$ & 32 & 2017 & 9 & diclofenac & 6.6 & 18 & 2 & 33 \\
$C_{\text{int}}$ & 28 & 2017 & 9 & diclofenac & 6.6 & 18 & 6 & 83 \\
$C_{\text{int}}$ & 29 & 2018 & 9 & diclofenac & 7.2 & 18 & 4 & 46 \\
$C_{\text{int}}$ & 12 & 2020 & 9 & diclofenac & 7.4 & 18 & 1 & 28 \\
$C_{\text{int}}$ & 33 & 2020 & 20 & diclofenac & 7.4 & 18 & 1 & 27 \\
$C_{\text{int}}$ & 3 & 2021 & 20 & diclofenac & 7.4 & 6 & 1 & 33 \\
$C_{\text{int}}$ & 19 & 2015 & 2 & diuron & 20.5 & 18 & 3 & 43 \\
$C_{\text{int}}$ & 20 & 2016 & 9 & diuron & 20.0 & 18 & 3 & 30 \\
$C_{\text{int}}$ & 20 & 2016 & 18 & diuron & 20.0 & 18 & 2 & 40 \\
$C_{\text{int}}$ & 30 & 2018 & 8 & diuron & 20.0 & 18 & 1 & 2 \\
$C_{\text{int}}$ & 30 & 2018 & 9 & diuron & 20.0 & 18 & 3 & 19 \\
$C_{\text{int}}$ & 14 & 2021 & 20 & diuron & 28.1 & 18 & 1 & 27 \\
$C_{\text{int}}$ & 23 & 2017 & 9 & naproxen & 135.0 & 18 & 3 & 42 \\
$C_{\text{int}}$ & 23 & 2017 & 9 & naproxen & 309.0 & 18 & 3 & 42 \\
$C_{\text{int}}$ & 24 & 2017 & 9 & naproxen & 135.0 & 18 & 2 & 25 \\
$C_{\text{int}}$ & 24 & 2017 & 9 & naproxen & 309.0 & 18 & 2 & 26 \\
$C_{\text{int}}$ & 26 & 2017 & 9 & naproxen & 135.0 & 18 & 4 & 38 \\
$C_{\text{int}}$ & 26 & 2017 & 9 & naproxen & 309.0 & 18 & 4 & 38 \\
$C_{\text{int}}$ & 27 & 2017 & 9 & naproxen & 135.0 & 18 & 6 & 34 \\
$C_{\text{int}}$ & 27 & 2017 & 9 & naproxen & 309.0 & 18 & 6 & 35 \\
$C_{\text{int}}$ & 13 & 2020 & 20 & naproxen & 307.0 & 18 & 1 & 28 \\
$C_{\text{int}}$ & 1 & 2021 & 12 & naproxen & 349.0 & 6 & 1 & 2 \\
$C_{\text{int}}$ & 1 & 2021 & 20 & naproxen & 349.0 & 6 & 1 & 26 \\
$C_{\text{int}}$ & 7 & 2022 & 12 & naproxen & 238.0 & 6 & 1 & 1 \\
$C_{\text{int}}$ & 7 & 2022 & 20 & naproxen & 238.0 & 6 & 1 & 23 \\
$C_{\text{int}}$ & 10 & 2022 & 20 & naproxen & 238.0 & 6 & 1 & 5 \\
\textit{Nrf2} & 34 & 2016 & 10 & diclofenac & 5.1 & 18 & 2 & 8 \\
\textit{Nrf2} & 34 & 2016 & 10 & diclofenac & 5.8 & 18 & 2 & 11 \\
\textit{Nrf2} & 34 & 2016 & 10 & diclofenac & 6.5 & 18 & 2 & 9 \\
\textit{Nrf2} & 34 & 2016 & 10 & diclofenac & 6.9 & 18 & 3 & 10 \\
\textit{Nrf2} & 34 & 2016 & 10 & diclofenac & 7.4 & 18 & 2 & 12 \\
\textit{Nrf2} & 35 & 2016 & 10 & diclofenac & 5.1 & 18 & 2 & 7 \\
\textit{Nrf2} & 35 & 2016 & 10 & diclofenac & 5.8 & 18 & 2 & 7 \\
\textit{Nrf2} & 35 & 2016 & 10 & diclofenac & 6.5 & 18 & 2 & 7 \\
\textit{Nrf2} & 35 & 2016 & 10 & diclofenac & 6.9 & 18 & 2 & 6 \\
\textit{Nrf2} & 35 & 2016 & 10 & diclofenac & 7.4 & 18 & 1 & 6 \\
\textit{Nrf2} & 36 & 2016 & 10 & diuron & 2.3 & 18 & 2 & 7 \\
\textit{Nrf2} & 36 & 2016 & 10 & diuron & 5.2 & 18 & 2 & 9 \\
\textit{Nrf2} & 36 & 2016 & 10 & diuron & 11.7 & 18 & 2 & 9 \\
\textit{Nrf2} & 36 & 2016 & 10 & diuron & 18.1 & 18 & 2 & 8 \\
\textit{Nrf2} & 36 & 2016 & 10 & diuron & 29.4 & 18 & 2 & 12 \\
\textit{Nrf2} & 36 & 2017 & 10 & naproxen & 135.0 & 18 & 2 & 8 \\
\textit{Nrf2} & 36 & 2017 & 10 & naproxen & 178.0 & 18 & 1 & 6 \\
\textit{Nrf2} & 36 & 2017 & 10 & naproxen & 234.0 & 18 & 2 & 8 \\
\textit{Nrf2} & 36 & 2017 & 10 & naproxen & 269.0 & 18 & 2 & 8 \\
\textit{Nrf2} & 36 & 2017 & 10 & naproxen & 309.0 & 18 & 2 & 11 \\
phenotpye & 40 & 2016 & 9 & diclofenac & 3.7 & 6 & 1 & 3 \\
phenotpye & 40 & 2016 & 9 & diclofenac & 4.6 & 6 & 1 & 3 \\
phenotpye & 40 & 2016 & 9 & diclofenac & 5.7 & 6 & 1 & 3 \\
phenotpye & 40 & 2016 & 9 & diclofenac & 7.2 & 6 & 1 & 3 \\
phenotpye & 40 & 2016 & 9 & diclofenac & 8.9 & 6 & 1 & 3 \\
phenotpye & 40 & 2016 & 9 & diclofenac & 11.2 & 6 & 1 & 3 \\
phenotpye & 40 & 2016 & 9 & diclofenac & 14.0 & 6 & 1 & 3 \\
phenotpye & 40 & 2016 & 9 & diclofenac & 17.5 & 6 & 1 & 3 \\
phenotpye & 40 & 2016 & 9 & diclofenac & 21.8 & 6 & 1 & 3 \\
phenotpye & 40 & 2016 & 9 & diclofenac & 27.3 & 6 & 1 & 3 \\
phenotpye & 40 & 2016 & 9 & diclofenac & 34.1 & 6 & 1 & 3 \\
phenotpye & 40 & 2016 & 9 & diclofenac & 42.6 & 6 & 1 & 3 \\
phenotpye & 40 & 2016 & 9 & diclofenac & 53.3 & 6 & 1 & 3 \\
phenotpye & 41 & 2016 & 9 & diclofenac & 3.2 & 6 & 1 & 4 \\
phenotpye & 41 & 2016 & 9 & diclofenac & 3.8 & 6 & 1 & 4 \\
phenotpye & 41 & 2016 & 9 & diclofenac & 4.5 & 6 & 1 & 4 \\
phenotpye & 41 & 2016 & 9 & diclofenac & 5.4 & 6 & 1 & 4 \\
phenotpye & 41 & 2016 & 9 & diclofenac & 6.5 & 6 & 1 & 4 \\
phenotpye & 41 & 2016 & 9 & diclofenac & 7.8 & 6 & 1 & 4 \\
phenotpye & 41 & 2016 & 9 & diclofenac & 9.4 & 6 & 1 & 4 \\
phenotpye & 41 & 2016 & 18 & diclofenac & 11.3 & 6 & 1 & 4 \\
phenotpye & 41 & 2016 & 9 & diclofenac & 13.5 & 6 & 1 & 4 \\
phenotpye & 41 & 2016 & 9 & diclofenac & 16.2 & 6 & 1 & 4 \\
phenotpye & 41 & 2016 & 9 & diclofenac & 19.5 & 6 & 1 & 4 \\
phenotpye & 41 & 2016 & 9 & diclofenac & 23.4 & 6 & 1 & 4 \\
phenotpye & 41 & 2016 & 9 & diclofenac & 28.1 & 6 & 1 & 4 \\
phenotpye & 43 & 2015 & 9 & diuron & 5.3 & 6 & 1 & 2 \\
phenotpye & 43 & 2015 & 9 & diuron & 6.4 & 6 & 1 & 2 \\
phenotpye & 43 & 2015 & 9 & diuron & 7.8 & 6 & 1 & 2 \\
phenotpye & 43 & 2015 & 9 & diuron & 9.3 & 6 & 1 & 2 \\
phenotpye & 43 & 2015 & 9 & diuron & 11.2 & 6 & 1 & 2 \\
phenotpye & 43 & 2015 & 9 & diuron & 13.5 & 6 & 1 & 2 \\
phenotpye & 43 & 2015 & 9 & diuron & 15.7 & 6 & 1 & 2 \\
phenotpye & 43 & 2015 & 9 & diuron & 19.4 & 6 & 1 & 2 \\
phenotpye & 43 & 2015 & 9 & diuron & 23.3 & 6 & 1 & 2 \\
phenotpye & 43 & 2015 & 9 & diuron & 27.9 & 6 & 1 & 2 \\
phenotpye & 43 & 2015 & 9 & diuron & 33.5 & 6 & 1 & 2 \\
phenotpye & 43 & 2015 & 9 & diuron & 40.2 & 6 & 1 & 2 \\
phenotpye & 43 & 2015 & 9 & diuron & 57.9 & 6 & 1 & 2 \\
phenotpye & 43 & 2015 & 9 & diuron & 69.5 & 6 & 1 & 2 \\
phenotpye & 43 & 2015 & 9 & diuron & 83.4 & 6 & 1 & 2 \\
phenotpye & 44 & 2015 & 9 & diuron & 2.1 & 6 & 1 & 2 \\
phenotpye & 44 & 2015 & 9 & diuron & 3.2 & 6 & 1 & 2 \\
phenotpye & 44 & 2015 & 9 & diuron & 4.8 & 6 & 1 & 2 \\
phenotpye & 44 & 2015 & 9 & diuron & 7.2 & 6 & 1 & 2 \\
phenotpye & 44 & 2015 & 9 & diuron & 8.5 & 6 & 1 & 2 \\
phenotpye & 44 & 2015 & 9 & diuron & 10.8 & 6 & 1 & 2 \\
phenotpye & 44 & 2015 & 9 & diuron & 12.8 & 6 & 1 & 2 \\
phenotpye & 44 & 2015 & 9 & diuron & 14.9 & 6 & 1 & 2 \\
phenotpye & 44 & 2015 & 9 & diuron & 16.1 & 6 & 1 & 2 \\
phenotpye & 44 & 2015 & 9 & diuron & 24.2 & 6 & 1 & 2 \\
phenotpye & 44 & 2015 & 9 & diuron & 36.3 & 6 & 1 & 2 \\
phenotpye & 44 & 2015 & 9 & diuron & 54.5 & 6 & 1 & 2 \\
phenotpye & 44 & 2015 & 9 & diuron & 81.7 & 6 & 1 & 2 \\
phenotpye & 47 & 2016 & 18 & naproxen & 10.6 & 6 & 1 & 4 \\
phenotpye & 47 & 2016 & 18 & naproxen & 21.2 & 6 & 1 & 4 \\
phenotpye & 47 & 2016 & 18 & naproxen & 42.3 & 6 & 1 & 4 \\
phenotpye & 47 & 2016 & 18 & naproxen & 84.7 & 6 & 1 & 4 \\
phenotpye & 47 & 2016 & 18 & naproxen & 169.0 & 6 & 1 & 4 \\
phenotpye & 47 & 2016 & 18 & naproxen & 339.0 & 6 & 1 & 4 \\
phenotpye & 47 & 2016 & 18 & naproxen & 677.0 & 6 & 1 & 4 \\
phenotpye & 47 & 2016 & 18 & naproxen & 1350.0 & 6 & 1 & 4 \\
phenotpye & 48 & 2016 & 18 & naproxen & 282.0 & 6 & 1 & 3 \\
phenotpye & 48 & 2016 & 18 & naproxen & 338.0 & 6 & 1 & 3 \\
phenotpye & 48 & 2016 & 18 & naproxen & 405.0 & 6 & 1 & 3 \\
phenotpye & 48 & 2016 & 18 & naproxen & 487.0 & 6 & 1 & 3 \\
phenotpye & 48 & 2016 & 18 & naproxen & 584.0 & 6 & 1 & 3 \\
phenotpye & 48 & 2016 & 18 & naproxen & 701.0 & 6 & 1 & 3 \\
phenotpye & 48 & 2016 & 18 & naproxen & 841.0 & 6 & 1 & 3 \\
phenotpye & 48 & 2016 & 18 & naproxen & 1010.0 & 6 & 1 & 3 \\
phenotpye & 48 & 2016 & 18 & naproxen & 1210.0 & 6 & 1 & 3 \\
phenotpye & 48 & 2016 & 18 & naproxen & 1450.0 & 6 & 1 & 3 \\
phenotpye & 49 & 2016 & 18 & naproxen & 137.0 & 6 & 1 & 3 \\
phenotpye & 49 & 2016 & 18 & naproxen & 165.0 & 6 & 1 & 3 \\
phenotpye & 49 & 2016 & 18 & naproxen & 198.0 & 6 & 1 & 3 \\
phenotpye & 49 & 2016 & 18 & naproxen & 237.0 & 6 & 1 & 3 \\
phenotpye & 49 & 2016 & 18 & naproxen & 285.0 & 6 & 1 & 3 \\
phenotpye & 49 & 2016 & 18 & naproxen & 342.0 & 6 & 1 & 3 \\
phenotpye & 49 & 2016 & 18 & naproxen & 410.0 & 6 & 1 & 3 \\
phenotpye & 49 & 2016 & 18 & naproxen & 492.0 & 6 & 1 & 3 \\
\end{longtable}

\pagebreak

\subsection{Parameter estimation}\label{si:sec:parameter_estimation}
 
\subsubsection{Bayesian parameter inference}\label{si:sec:bayesian-parameter-inference}

Working with temporally resolved exposure and 'omics data means integrating different datasets from numerous biological experiments into the modeling and optimization/inference process.
This is associated with considerable experimental noise, which complicates the process of obtaining accurate parameters that determine the model. Accounting for measurement error and biological stochasticity (e.g.~in lethality or sublethal effects) in the data can help to infer the true effect of a chemical by separating noise from effect.
Bayesian parameter inference approaches accommodate all these necessities and, further, report parameter distributions, reflecting the uncertainty in the true parameter and also provide estimates of the expected variation in experimental observations.
These so called, posterior predictions, are excellent tools for gauging the predictive capacities of the model and on top of that provide an ideal validation tool for novel incoming data.
At the heart of the Bayesian philosphy is a process that is called bayesian updating.
In order to explain this a short excurse into the Bayes rule is necessary. 

\begin{align*}
   Pr(\theta~|~Y) &= \frac{Pr(\theta)~Pr(Y~|~\theta)}{Pr(Y)}  \numberthis \label{si:eq:conditional_probability} \\
   Posterior &\approx Prior \times Likelihood  \numberthis \label{si:eq:posterior}
\end{align*}

Equation \ref{si:eq:conditional_probability} is also known as Bayes Theorem and it is used to calculate conditional probabilities.
It reads as: The probability of a set of parameters $\theta$, conditional on the observed data $Y$ is equal to the joint probability of the parameters and the probability of the likelihood of the data given the parameters (and the model to relate parameters to data), divided by the probability of the observations $Y$.
Because the calculation of the denominator $Pr(Y)$ of the equation is complicated, it is usually ignored due to its independence of the parameters (no $\theta$ is involved) and considered as a proportionality constant, which ensures that the resulting probability function integrates to 1.
Thus, Equation \ref{si:eq:posterior} contains the remaining components of Equation \ref{si:eq:conditional_probability} that actually bear relevant concepts for the understanding of uncertainty in statistics---$posterior$, $prior$ and $likelihood$.
The notation of the likelihood in Equation \ref{si:eq:conditional_probability} is a convention. 
One could also rewrite the likelihood as $Pr(Y~|~f(\theta))$ to more explicitly express that the likelihood of the data not only depends on the parameter values but also on the used mathematical model $f$, that describes how the parameters $\theta$ are transformed to predict the observed data $Y$.
To be clear, in the scope of this work $f$ is a TKTD model. For more information on the theory of bayesian inference, we refer to the excellent handbook \emph{Statistical Rethinking} \cite{McElreath.2015}.

\subsubsection{Leveraging modern probabilistic programming languages (PPL) to solve the computational challenges of the 'omics integration into TKTD models}

The difficulty of performing Bayesian parameter inference on an ODE model is that systems of ODEs need to be solved 10,000--100,000s of times in order to first converge on the typical set of parameters, and the sample frequently enough to build a valid approximation of the posterior parameter distribution.
Such a procedure may take a long time, if the ODE system is solved in each iteration.
Fortunately, modern probabilistic programming languages (PPL) provide the tools, to address such difficulties very efficiently, by exploiting auto differentiation with an \emph{adjoint sensitivity} approach and by applying compilation to highly efficient symbolic languages.
In this work, \textit{numypro} \cite{Bingham.2018,Phan.2019} was used as a PPL (other highly recommendable choices are Stan, or pymc). 
Numpyro uses \textit{JAX} \cite{Bradbury.2018} to compile ODE models and deliver solutions along with autodifferentiated gradients, with respect to the model parameters. 
By using JAX, the 202 ODE systems needed to integrate all datasets into one model could be evaluated very efficiently resulting in a model evaluation time of 40 ms for 1 iteration.
Still a computational problem remains, because for using the state of the art MCMC method, NUTS \cite{Hoffman.2011}, the likelihood function (and its gradients) need to be computed for each data point.
In the given dataset, this means 1426 gradient evaluations with respect to all model parameters per leapfrog step (the number of leapfrosteps varied between 1--1023 per iteration).
This easily scales to dimensions where gradient based MCMC approaches, like NUTS have difficulties, especially when the ODE model and therefore the likelihood function and its gradients, becomes more complex.
For simple problem like the 4-parameter GUTS model $k_d$, $k_k$, $h_b$, $z$, solving the problem with a NUTS approach is feasible (walltime $\approx 30$ minutes), but with more complex models with higher number of parameters, NUTS approaches quickly becomes infeasible (walltime \textgreater{} 48 h). 
In these situation, posteriors were approximated with stochastic variational inference (SVI) \cite{Blei.2017}, which estimates posterior distributions, based on finding a parametric distribution that approximates the true, unknown posterior distribution.
While these methods, are constrained to deliver parametric posteriors, they were in good agreement with the posteriors produced by the NUTS algorithm.
In order to address the prerequisites of highly fragmented and complex datasets for bayesian parameter inference, \texttt{pymob} (\url{https://github.com/flo-schu/pymob}) is being developed as a modeling framework, which allows the user to switch between inference frameworks (e.g.~interactive, maximum-likelihood, approximate-bayes, fully-bayesian), while maintaining a consistent deterministic and stochastic model formulation.
In essence to minimize the frustrating dimensional overhead when working with complex datasets and seamlessly switch between different tools, which may work better or worse with different demands of the data and model.

In order to assess parameter uncertainty and identify potential identifiability issues, 100 markov chains (NUTS) or 100 SVI approximations, were started with initial parameters drawn from a uniform interval from -1 to 1, which were subsequently transformed to the scales priors of the parameter distributions.
This effectively detects any local minimima with likelihoods very close to the global optimum and thus indicate the presence of parameter identifiability issues.
The exact algorithm is described in \Cref{si:sec:parameter_analysis_algorithm}.

\subsubsection{Parameter analysis algorithm}\label{si:sec:parameter_analysis_algorithm}

The parameter uncertainty analysis algorithm follows these steps: 
\begin{enumerate}
    \item Start 100 parameter estimations with SVI with a multivariate normal guide that maps onto log-normal distributions, containing the prior information. The starting values are drawn from a uniform distribution between {[}-1, 1{]}, which will cover most of the majority of the prior density. The interval is not selected larger, because already this range contains parameter estimates that impose severe difficulties for the solver. SVI is parameterized with a learning rate of 0.001 and 50,000 iterations. Usually convergence happens within the first few 1000 iterations. Updates of the estimator that contain infeasible values (inf, nan; occur, when the solver reaches max. stepsize because of occasional extreme parameter combinations). 
    \item After 4 hours, all running estimations are terminated, because in all probability they were stuck on some very slow local minimum, where evaluations of the detemrinistic solver takes a long time. The 100 chains are combined into one large posterior and saved. 
    \item Following this, the chains are clustered, by comparing whether the means are within 1 standard deviation of any other chain. Alternatively, a fraction of the mean can be given as a deviation criterion Next, parameter analysis are conducted for each cluster (Filtered to those estimates that have a log-likelihood less negative than 1.1 times the least negative likelihood) 
    \item Calculate BIC write a table (Tex) of the parameters and their standard deviations. 
    Plot log-histograms of the parameters Plot joint distributions of pairwise parameters to visually inspect clusters Plot posterior predictions of the different substances These estimates are used solely for model improvement and understanding the different stable modes in the case of parameter-identifiability issues. 
    \item Finally, the same procedure as described in (4) is conducted for all estimates with a likelihood deviation of 1.025 (2.5\%) below the estimate with the highest likelihood. 
\end{enumerate}

This estimates within 2.5\% variation of the best estimtate are used for reporting in the study. It should reflect the uncertainty very well. In case not many chains are in this estimate, it indicates that the number of iterations were too few, or the learning rate too low, or the prior distributions too wide. Ideally estimation should be repeated until a reasonable number of independent estimates (\textgreater{} 5) have converged on the same maximal likelihood. Of course this is still no guarantee that the global best estimate has been found, but it is a good indication.
\pagebreak

\subsection{GUTS-RNA-pulse model}

\subsubsection{Model description of the GUTS-RNA-pulse model (substance specific and substance independent)}\label{si:mod:guts-rna-pulse}

\begin{align}
\frac{dC_i}{dt} &= k_i~C_e - k_m~C_i~P \\
\frac{dR}{dt} &= r_{rt}~\text{activation}(C_i,~C_{i,max},~ z_{ci},~ v_{rt}) - k_{rd} ~ (R - R_0) \\
\frac{dP}{dt} &= k_p~ ((R - R_0) - P) \\
h(t) &= k_k~ max(0, R(t) - z) +  h_b \\
S(t) &= e^{-\int_0^t h(t) dt}
\end{align}

with 

\begin{equation}
    activation(C_i,~C_{i,max},~z_{ci}, ~v_{rt}) = 0.5 + \frac{1}{\pi} ~ arctan(v_{rt} ~ (\frac{C_i}{C_{i,max}} - z_{ci}))
\end{equation}

\begin{table}[h!]
    \centering
    \footnotesize
    \begin{threeparttable}
        \caption{TKTD Parameters used in the GUTS-RNA-pulse model. The column ``Assumed substance independence'' indicates whether a parameter is supposed to be shared for multiple  substances.}
        \label{si:tab:parameters}
        \begin{tabularx}{\textwidth}{s X s s}
        \toprule
        Parameter              & Definition & Unit & Assumed substance independence \\
        \midrule
        ${k}_{i}$              & Uptake rate constant of the chemical into the internal compartment of the ZFE & $h^{-1}$ & no \\
        ${k}_{m}$              & Metabolization rate constant from the internal compartment of the ZFE& $\frac{L}{\mu mol~h}$ & no \\
        ${z}_{\text{ci}}$      & Scaled internal concentration threshold for the activation of \textit{nrf2} expression & $\frac{\mu mol~L^{-1}}{\mu mol~L^{-1}}$ & no \\
        ${v}_{\text{rt}}$      & Scaled responsiveness of the \textit{nrf2} activation (slope of the activation function) & $\frac{\mu mol~L^{-1}}{\mu mol~L^{-1}}$ & yes/no \tnote{\textit{a}} \\
        ${r}_{\text{rt}}$      & Constant \textit{nrf2} expression rate after activation \tnote{\textit{b}} & fc \tnote{\textit{c}} & yes \\
        ${k}_{\text{rd}}$      & Nrf2 decay rate constant & $h^{-1}$ & yes \\
        ${k}_{p}$              & Dominant rate constant of synthesis and decay of metabolizing proteins & $h^{-1}$ & yes \\
        ${z}$                  & Effect \textit{nrf2}-threshold of the hazard function \tnote{\textit{b}} & fc \tnote{\textit{c}} & yes \\
        ${k}_{k}$              & killing rate constant for \textit{nrf2} \tnote{\textit{b}} & $fc^{-1}~h^{-1}$ \tnote{\textit{c}} & yes\\
        ${h}_{b}$              & background hazard rate constant & $h^{-1}$ & yes \\
        $\sigma_{\text{cint}}$ & Log-normal error of the internal concentration & & yes \\
        $\sigma_{nrf2}$ & Log-normal error of the \textit{nrf2} expression \tnote{\textit{b}} & & yes \\
        \bottomrule
        \end{tabularx}    
        \begin{tablenotes}
            \item[\textit{a}] In an unscaled version of the activation function, $v_{rt}$ is not considered substance independent, due to an inverse relationship between $v_{rt}$ and $C_{i,max}$
            \item[\textit{b}] relative to the \textit{nrf2} concentration in untreated ZFE (fold-change)
            \item[\textit{c}] fold change: $\frac{\mu mol~nrf2\text{-treatment}~L^{-1}}{\mu mol~nrf2\text{-control}~L^{-1}}$ 
        \end{tablenotes}
    \end{threeparttable}
\end{table}

\pagebreak
\subsubsection{Model fits for GUTS-RNA-pulse}\label{si:sec:model_fits_GUTS_RNA_pulse}

\begin{center}
    \includegraphics[height=.27\textheight]{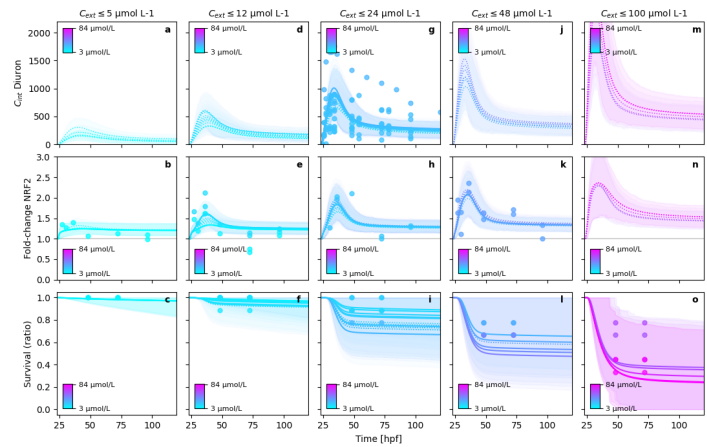}
    \captionof{figure}{Posterior predictions of the GUTS-RNA-pulse model for diuron }
    \label{si:fig:model_fits_guts_rna_pulse_specific_diuron}
\end{center}

\begin{center}
    \includegraphics[height=.27\textheight]{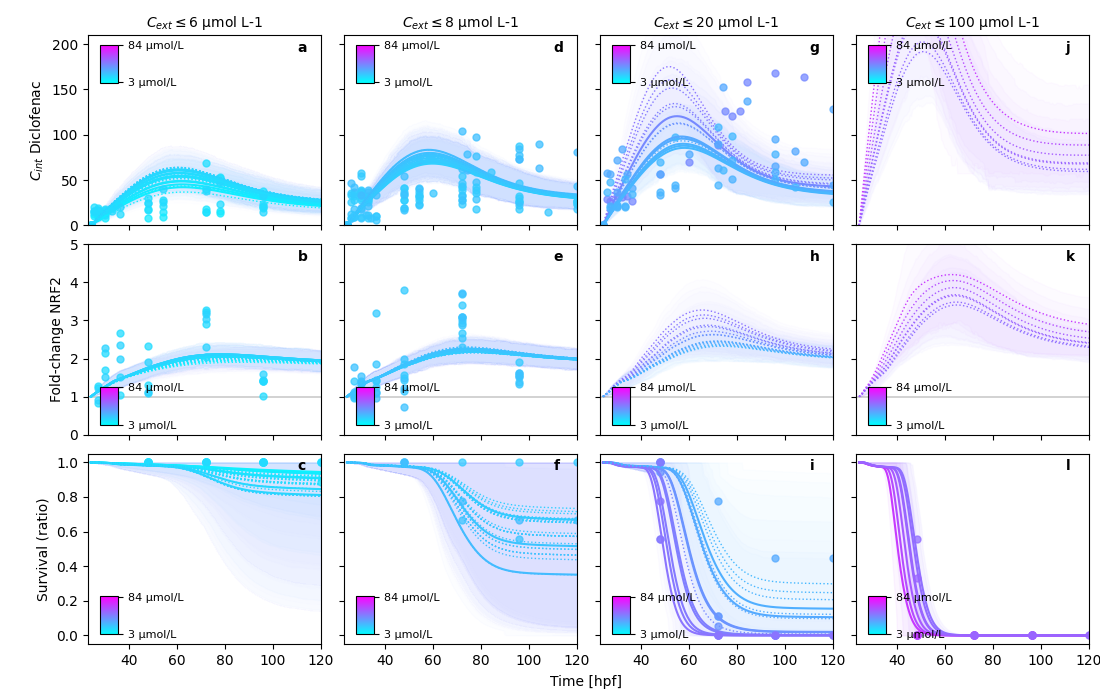}
    \captionof{figure}{Posterior predictions of the GUTS-RNA-pulse model for diclofenac }
    \label{si:fig:model_fits_guts_rna_pulse_specific_diclofenac}
\end{center}

\begin{center}
    \includegraphics[height=.27\textheight]{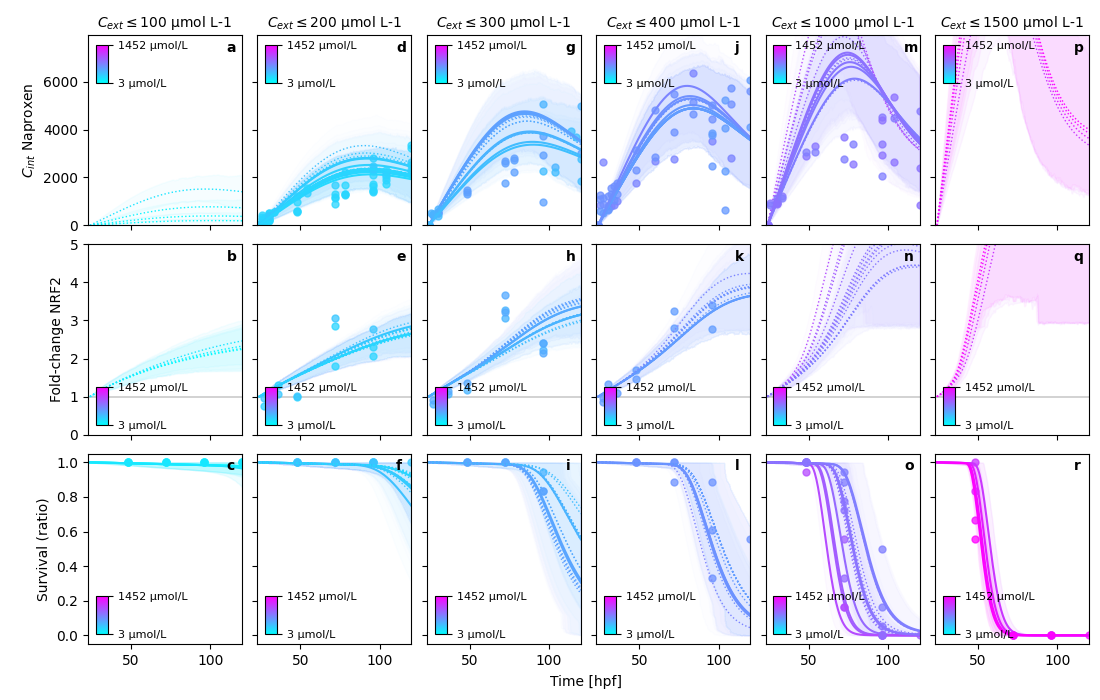}
    \captionof{figure}{Posterior predictions of the GUTS-RNA-pulse model for naproxen }
    \label{si:fig:model_fits_guts_rna_pulse_specific_naproxen}
\end{center}

\begin{center}
    \centering
    \includegraphics[height=0.9\textheight]{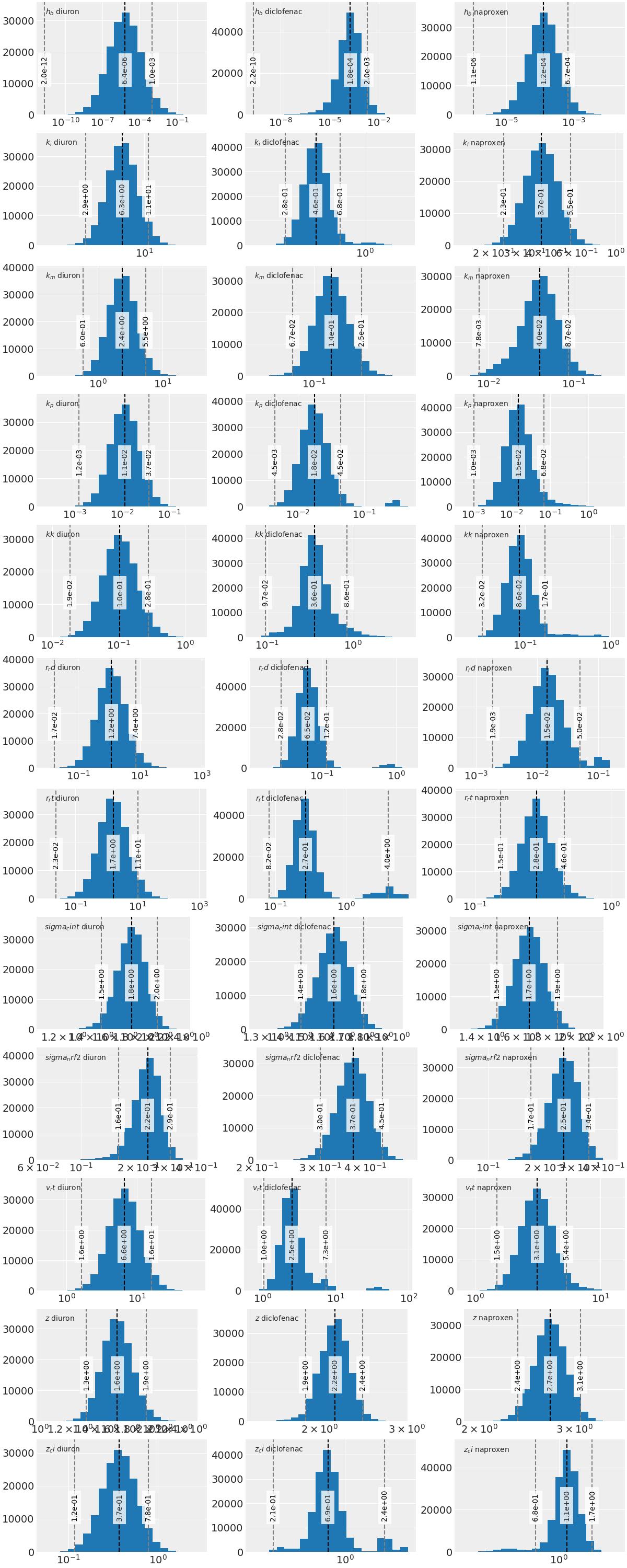}
    \captionof{figure}{Parameter estimates of the GUTS-RNA-pulse model with substance specific parameters.}
    \label{si:fig:parameter_estimates_hist_rna_pulse_3_6c_substance_specific}
\end{center}

\begin{minipage}\linewidth
\captionof{table}{Parameter estimates and posterior highest densitiy intervals (HDI) of the substance specific GUTS-RNA-pulse model. The HDI contains 94\% of the probable parameter values given the data.}
\label{si:tab:parameters-guts-rna-pulse}
\begin{tabular}{llllllllll}
\toprule
Parameters & \multicolumn{3}{r}{Diuron} & \multicolumn{3}{r}{Diclofenac} & \multicolumn{3}{r}{Naproxen} \\
 & mean & hdi 3\% & hdi 97\% & mean & hdi 3\% & hdi 97\% & mean & hdi 3\% & hdi 97\% \\
\midrule
${k}_{i}$ & 6.62 & 3.06 & 10.69 & 0.48 & 0.29 & 0.66 & 0.38 & 0.24 & 0.54 \\
${k}_{m}$ & 2.71 & 0.69 & 5.30 & 0.15 & 0.07 & 0.24 & 0.04 & 0.01 & 0.08 \\
${z}_{\text{ci}}$ & 0.41 & 0.12 & 0.75 & 0.84 & 0.21 & 2.28 & 1.14 & 0.72 & 1.70 \\
${v}_{\text{rt}}$ & 7.60 & 1.77 & 15.30 & 3.63 & 1.03 & 6.55 & 3.32 & 1.53 & 5.19 \\
${r}_{\text{rt}}$ & 3.28 & 0.04 & 9.64 & 0.65 & 0.09 & 3.63 & 0.30 & 0.15 & 0.44 \\
${k}_{\text{rd}}$ & 2.28 & 0.02 & 6.60 & 0.08 & 0.03 & 0.11 & 0.02 & 0.00 & 0.04 \\
${k}_{p}$ & 0.01 & 0.00 & 0.03 & 0.03 & 0.00 & 0.04 & 0.03 & 0.00 & 0.06 \\
${z}$ & 1.61 & 1.34 & 1.90 & 2.15 & 1.89 & 2.43 & 2.71 & 2.36 & 3.06 \\
${k}_{k}$ & 0.12 & 0.02 & 0.27 & 0.41 & 0.09 & 0.79 & 0.10 & 0.03 & 0.16 \\
${h}_{b}$ & 0.00 & 0.00 & 0.00 & 0.00 & 0.00 & 0.00 & 0.00 & 0.00 & 0.00 \\
${\sigma}_{\text{cint}}$ & 1.76 & 1.49 & 2.01 & 1.61 & 1.46 & 1.76 & 1.67 & 1.49 & 1.86 \\
${\sigma}_{\text{nrf2}}$ & 0.22 & 0.16 & 0.29 & 0.37 & 0.30 & 0.44 & 0.25 & 0.17 & 0.33 \\
\bottomrule
\end{tabular}
\end{minipage}

\pagebreak
\subsubsection{Model fits for GUTS-RNA-pulse model with parameter sharing for the RNA and protein modules}

\begin{center}
    \includegraphics[height=.27\textheight]{assets/fig_si_model_fits_rna_pulse_3_6c_substance_independent_rna_protein_module_diuron.png}
    \captionof{figure}{Posterior predictions of diuron for the parameter sharing GUTS-RNA-pulse model.}
    \label{si:fig:model_fits_guts_rna_pulse_independent_diuron}
\end{center}

\begin{center}
    \includegraphics[height=.27\textheight]{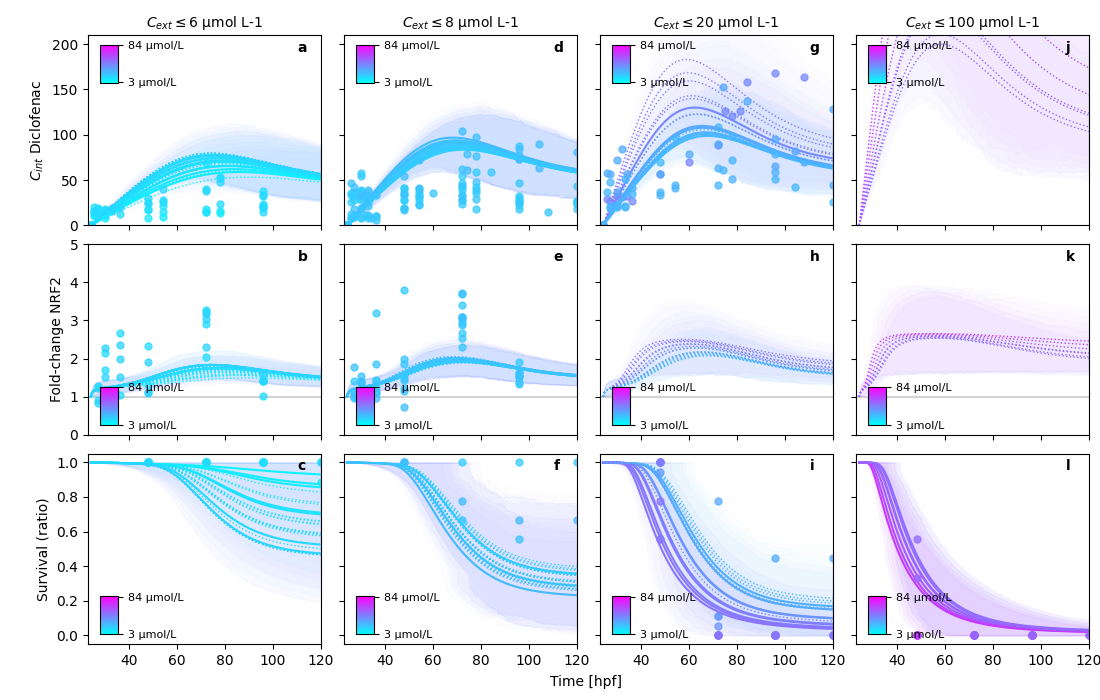}
    \captionof{figure}{Posterior predictions of diclofenac for the parameter sharing GUTS-RNA-pulse model.}
    \label{si:fig:model_fits_guts_rna_pulse_independent_diclofenac}
\end{center}

\begin{center}
    \includegraphics[height=.27\textheight]{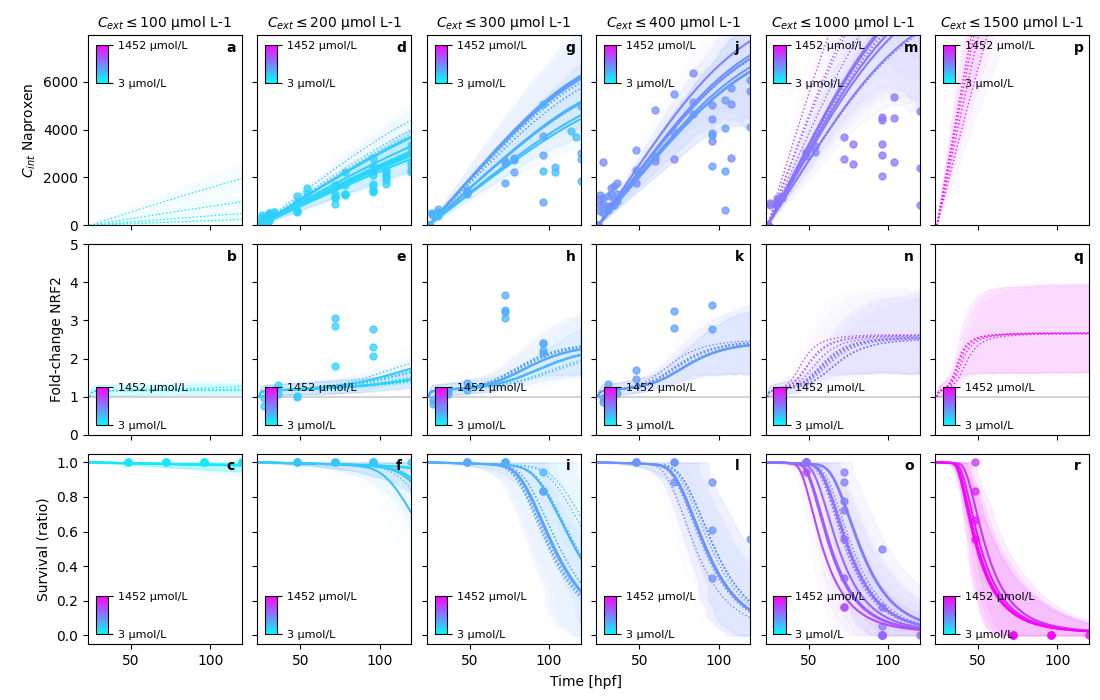}
    \captionof{figure}{Posterior predictions of naproxen for the parameter sharing GUTS-RNA-pulse model.}
    \label{si:fig:model_fits_guts_rna_pulse_independent_naproxen}
\end{center}

\begin{center}
    \centering
    \includegraphics[height=0.9\textheight]{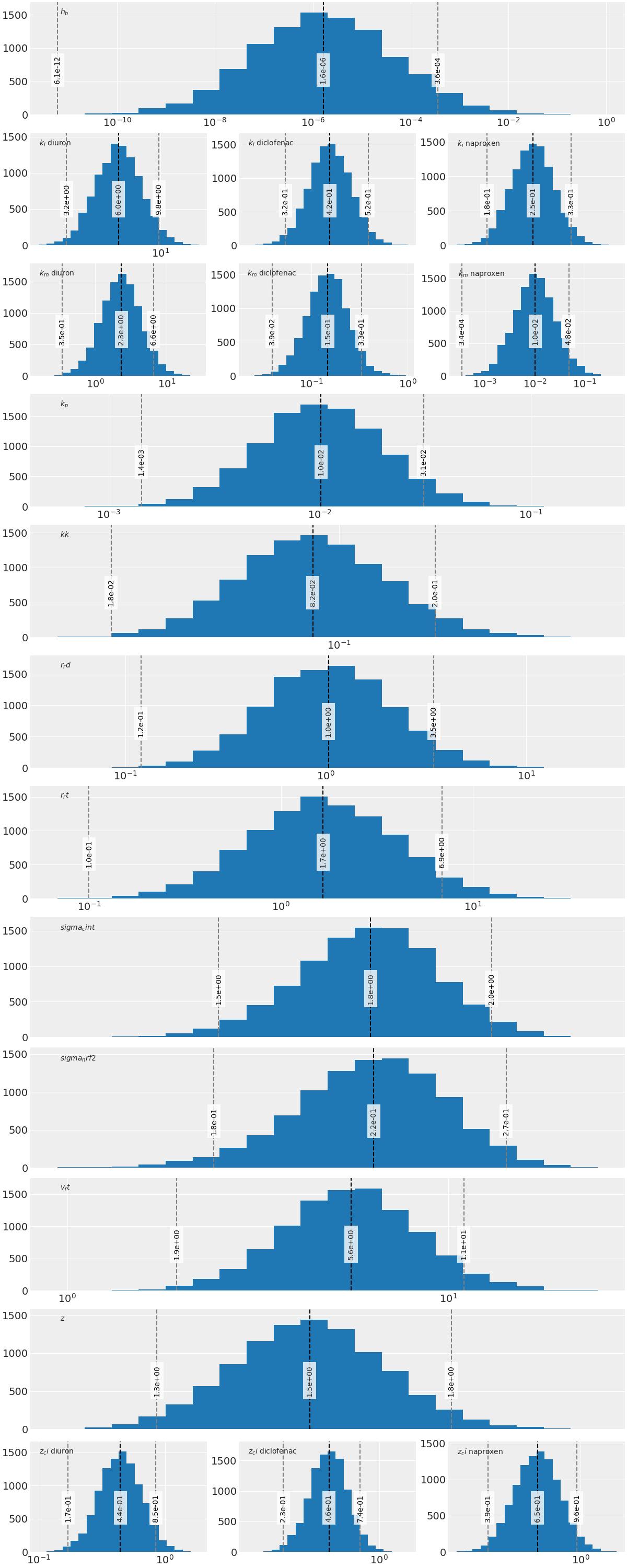}
    \captionof{figure}{Parameter estimates of the GUTS-RNA-pulse model with substance-independent parameters for the RNA and protein dynamics.}
    \label{si:fig:parameter_estimates_hist_rna_pulse_3_6c_substance_independent}
\end{center}

\begin{minipage}{\linewidth}
\centering
\captionof{table}{Parameter estimates and posterior highest densitiy intervals (HDI) of the GUTS-RNA-pulse model with a substance independent RNA protein module. Parameters which share information between substances are given in the form (mean (3\% hdi--97\% hdi)). Parameter sharing reduces the number of parameters for all 3 substances from 30 to 18.}
\label{si:tab:parameters-rna-pulse-independent}
\begin{tabularx}{\textwidth}{X X X X X X X X X X}
    \toprule
     & \multicolumn{3}{c}{diuron} & \multicolumn{3}{c}{diclofenac} & \multicolumn{3}{c}{naproxen} \\
     & mean & hdi 3\% & hdi 97\% & mean & hdi 3\% & hdi 97\% & mean & hdi 3\% & hdi 97\% \\
    \midrule
    $k_i$ & 6.21 & 3.29 & 9.67 & 0.42 & 0.33 & 0.52 & 0.25 & 0.19 & 0.33 \\
    $k_m$ & 2.83 & 0.41 & 6.26 & 0.17 & 0.04 & 0.32 & 0.02 & 0.00 & 0.04 \\
    $z_{ci}$ & 0.48 & 0.17 & 0.82 & 0.47 & 0.24 & 0.73 & 0.66 & 0.39 & 0.94 \\
    $v_{rt}$ & \multicolumn{9}{c}{6.02 (2.13--10.76)} \\
    $r_{rt}$ & \multicolumn{9}{c}{2.42 (0.08--6.34)} \\
    $k_{rd}$ & \multicolumn{9}{c}{1.35 (0.13--3.27)} \\
    $k_p$    & \multicolumn{9}{c}{0.01 (0.00--0.03)} \\
    $z$      & \multicolumn{9}{c}{1.55 (1.30--1.80)} \\
    $k_k$    & \multicolumn{9}{c}{0.10 (0.02--0.20)} \\
    $h_b$    & \multicolumn{9}{c}{0.00 (0.00--0.00)} \\
    $\sigma_{cint}$ & \multicolumn{9}{c}{1.76 (1.52--1.99)} \\
    $\sigma_{Nrf2}$ & \multicolumn{9}{c}{0.22 (0.18--0.27)} \\
    \bottomrule
\end{tabularx}
\end{minipage}
    
\pagebreak
\subsection{GUTS-RNA model}

\subsubsection{Model description of the GUTS-RNA model}\label{si:mod:guts-rna}

\begin{align}
\frac{dC_i}{dt} &= k_i~C_e - k_e~C_i \\
\frac{dD}{dt} &= k_a~C_i - k_r~D \\
h(t) &= k_k~ max(0, D(t) - z) +  h_b \\
S(t) &= e^{-\int_0^t h(t) dt}
\end{align}

\begin{table}[h!]
    \centering
    \footnotesize
    \begin{threeparttable}
        \caption{TKTD Parameters used in the GUTS-RNA model.}
        \label{si:tab:parameters-guts-rna}
        \begin{tabularx}{\textwidth}{s X s}
        \toprule
        Parameter              & Definition & Unit\\
        \midrule
        ${k}_{i}$              & Uptake rate constant of the chemical into the internal compartment of the ZFE & $h^{-1}$ \\
        ${k}_{e}$              & Elimination rate constant from the internal compartment of the ZFE & $h^{-1}$ \\
        ${k}_{a}$              & Damage accrual rate constant & $h^{-1}$ \\
        ${k}_{r}$              & Damage repair rate constant & $h^{-1}$ \\
        ${z}$                  & Effect \textit{nrf2}-threshold of the hazard function \tnote{\textit{a}} & fc \tnote{\textit{b}}  \\
        ${k}_{k}$              & killing rate constant for \textit{nrf2} \tnote{\textit{a}} & $fc^{-1}~h^{-1}$ \tnote{\textit{b}} \\
        ${h}_{b}$              & background hazard rate constant & $h^{-1}$ \\
        $\sigma_{\text{cint}}$ & Log-normal error of the internal concentration &  \\
        $\sigma_{nrf2}$        & Log-normal error of the \textit{nrf2} expression \tnote{\textit{a}} &  \\
        \bottomrule
        \end{tabularx}    
        \begin{tablenotes}
            \item[\textit{a}] relative to the \textit{nrf2} concentration in untreated ZFE (fold-change)
            \item[\textit{b}] fold change: $\frac{\mu mol~nrf2\text{-treatment}~L^{-1}}{\mu mol~nrf2\text{-control}~L^{-1}}$ 
        \end{tablenotes}
    \end{threeparttable}
\end{table}

\pagebreak

\pagebreak
\subsubsection{Model fits for GUTS-RNA}\label{sec:si.model_fits_guts_RNA}

\begin{center}
    \includegraphics[height=.27\textheight]{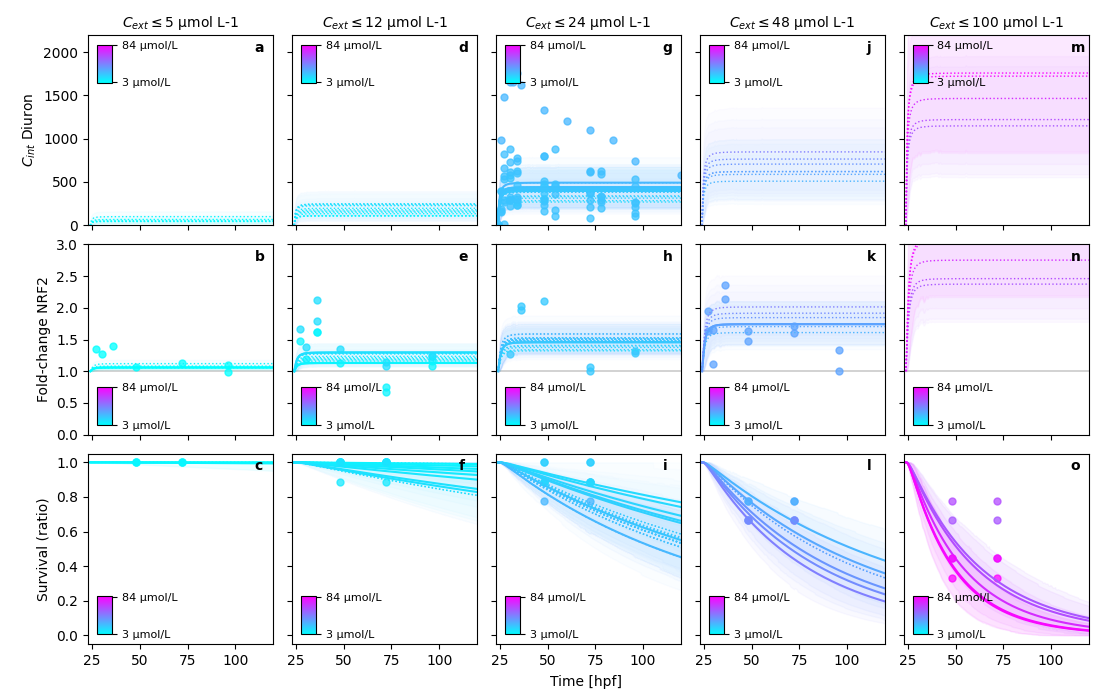}
    \captionof{figure}{Posterior predictions of diuron for GUTS-RNA model.}
    \label{si:fig:model_fits_guts_rna_diuron}
\end{center}

\begin{center}
    \includegraphics[height=.27\textheight]{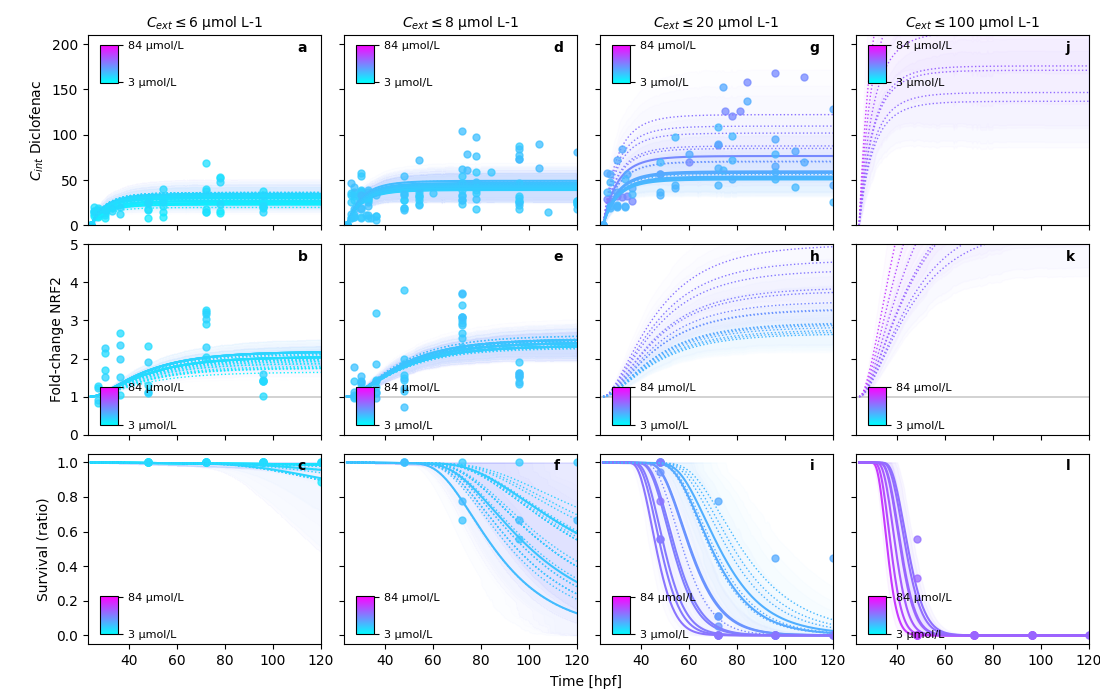}
    \captionof{figure}{Posterior predictions of diclofenac for GUTS-RNA model.}
    \label{si:fig:model_fits_guts_rna_diclofenac}
\end{center}

\begin{center}
    \includegraphics[height=.27\textheight]{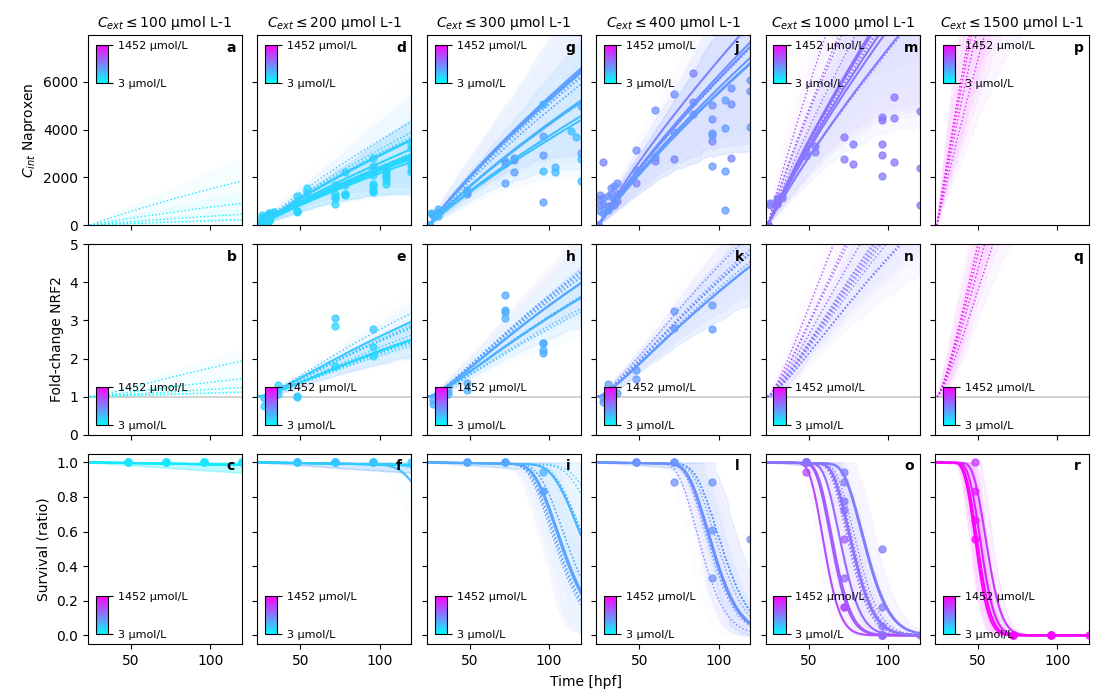}
    \captionof{figure}{Posterior predictions of naproxen for GUTS-RNA model.}
    \label{si:fig:model_fits_guts_rna_naproxen}
\end{center}

\begin{center}
    \centering
    \includegraphics[height=0.9\textheight]{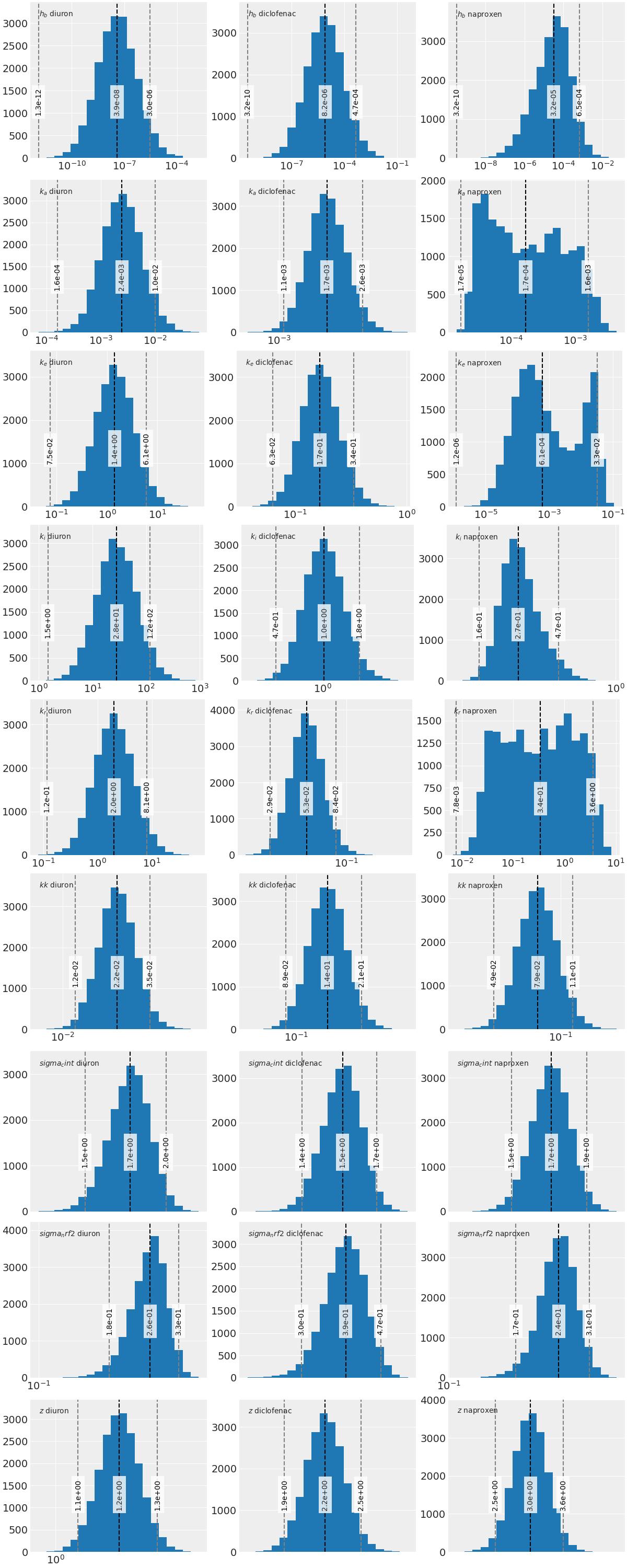}
    \captionof{figure}{Parameter estimates of the GUTS-RNA model.}
    \label{si:fig:parameter_estimates_guts_rna}
\end{center}

\begin{minipage}{\linewidth}
\centering
\captionof{table}{Parameter estimates and posterior highest density intervals (HDI) of the GUTS-RNA model. The HDI contains 94\% of the probable parameter values given the data.}
\label{tab:parameters-guts__guts_full_rna}
\begin{tabular}{llllllllll}
\toprule
Parameters & \multicolumn{3}{r}{Diuron} & \multicolumn{3}{r}{Diclofenac} & \multicolumn{3}{r}{Naproxen} \\
 & mean & hdi 3\% & hdi 97\% & mean & hdi 3\% & hdi 97\% & mean & hdi 3\% & hdi 97\% \\
\midrule
${k}_{i}$ & 41.74 & 1.48 & 110.75 & 1.08 & 0.49 & 1.74 & 0.29 & 0.16 & 0.45 \\
${k}_{e}$ & 2.11 & 0.08 & 5.61 & 0.18 & 0.07 & 0.32 & 0.01 & 0.00 & 0.03 \\
${k}_{a}$ & 0.00 & 0.00 & 0.01 & 0.00 & 0.00 & 0.00 & 0.00 & 0.00 & 0.00 \\
${k}_{r}$ & 2.87 & 0.10 & 7.47 & 0.05 & 0.03 & 0.08 & 0.89 & 0.01 & 3.33 \\
${z}$ & 1.19 & 1.07 & 1.31 & 2.19 & 1.89 & 2.47 & 3.03 & 2.55 & 3.53 \\
${k}_{k}$ & 0.02 & 0.01 & 0.03 & 0.14 & 0.09 & 0.20 & 0.08 & 0.05 & 0.11 \\
${h}_{b}$ & 0.00 & 0.00 & 0.00 & 0.00 & 0.00 & 0.00 & 0.00 & 0.00 & 0.00 \\
${\sigma}_{\text{cint}}$ & 1.74 & 1.48 & 1.99 & 1.53 & 1.38 & 1.68 & 1.72 & 1.53 & 1.91 \\
${\sigma}_{\text{nrf2}}$ & 0.26 & 0.19 & 0.33 & 0.39 & 0.31 & 0.47 & 0.24 & 0.18 & 0.31 \\
\bottomrule
\end{tabular}
\end{minipage}

\pagebreak
\subsection{GUTS-scaled-damage model}

\subsubsection{Model description of the GUTS-scaled-damage model}\label{si:mod:guts-scaled-damage}

\begin{align}
\frac{dC_i}{dt} &= k_i~C_e - k_e~C_i \\
\frac{dD}{dt} &= k_d \cdot (C_i-D) \\
h(t) &= k_k~ max(0, D(t) - z) +  h_b \\
S(t) &= e^{-\int_0^t h(t) dt}
\end{align}

\begin{table}[h!]
    \centering
    \footnotesize
        \caption{TKTD Parameters used in the GUTS-scaled-damage model.}
        \label{si:tab:parameters-guts-scaled-damage}
        \begin{tabularx}{\textwidth}{s X s}
        \toprule
        Parameter              & Definition & Unit\\
        \midrule
        ${k}_{i}$              & Uptake rate constant of the chemical into the internal compartment of the ZFE & $h^{-1}$ \\
        ${k}_{e}$              & Elimination rate constant from the internal compartment of the ZFE & $h^{-1}$ \\
        ${k}_{d}$              & Dominant rate constant of damage dynamics & $h^{-1}$ \\
        ${z}$                  & Effect damage-threshold of the hazard function  & $\mu mol~L^{-1}$ \\
        ${k}_{k}$              & killing rate constant & $L~\mu mol^{-1}~h^{-1}$  \\
        ${h}_{b}$              & background hazard rate constant & $h^{-1}$ \\
        $\sigma_{\text{cint}}$ & Log-normal error of the internal concentration &  \\
        \bottomrule
        \end{tabularx}    
\end{table}

\pagebreak

\pagebreak
\subsubsection{Model fits for GUTS-scaled-damage}\label{si:sec:model_fits_guts_scaled_damage}

\begin{center}
    \includegraphics[height=.27\textheight]{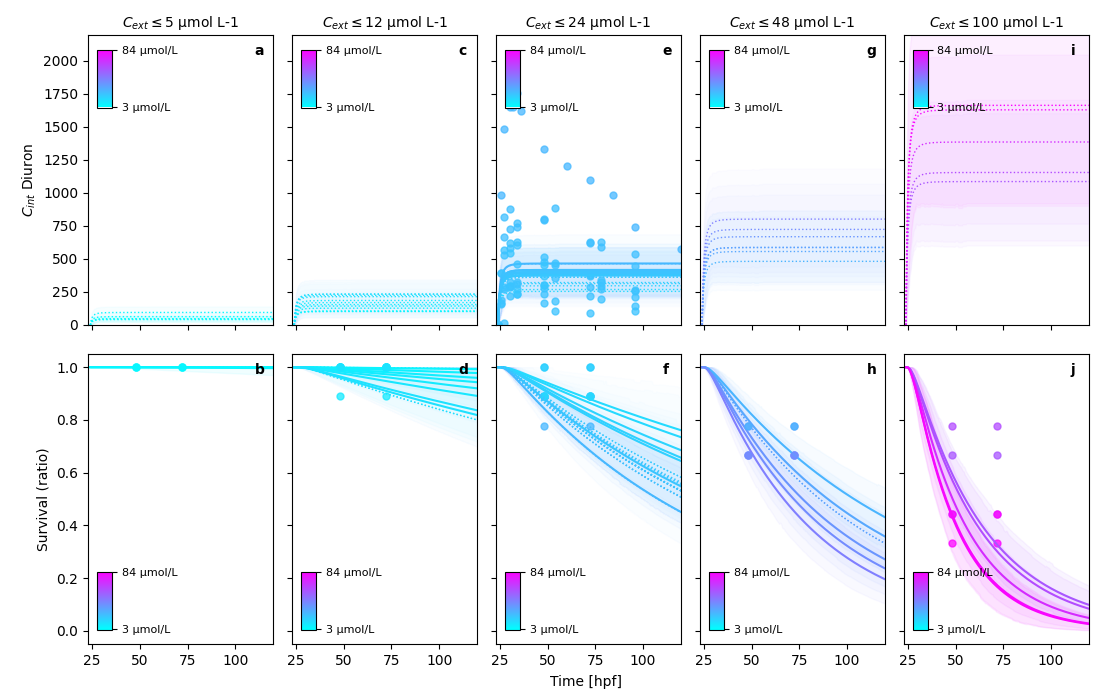}
    \captionof{figure}{Posterior predictions of diuron for GUTS-scaled damage model.}
    \label{fig:si.model_fits_guts_scaled_damage_diuron}
\end{center}

\begin{center}
    \includegraphics[height=.27\textheight]{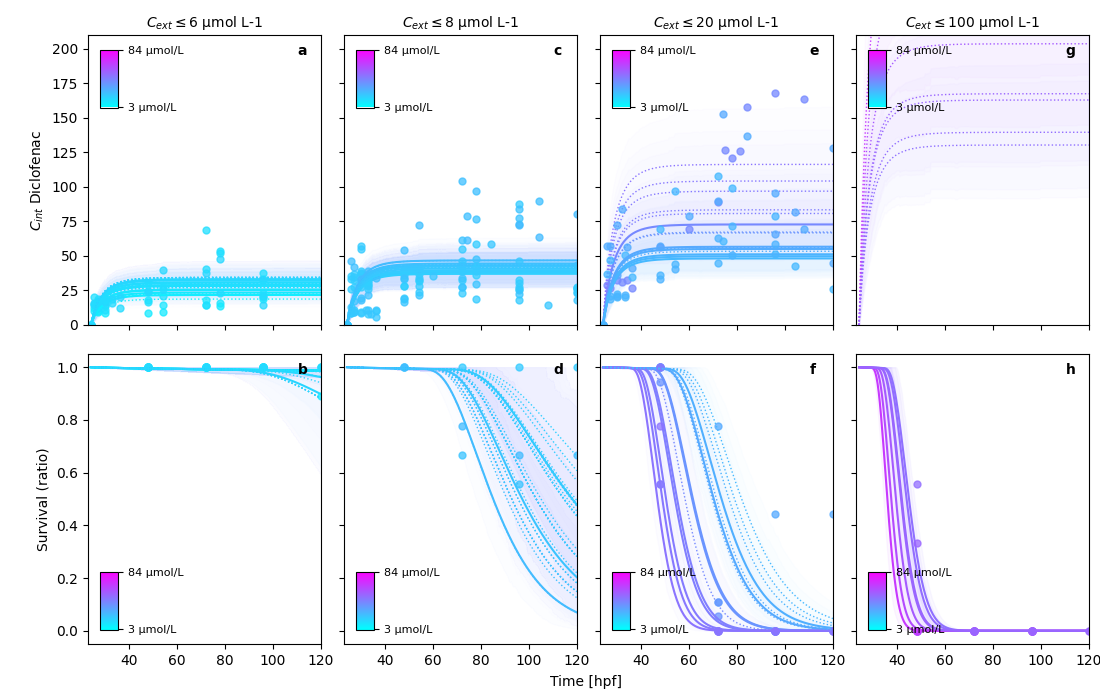}
    \captionof{figure}{Posterior predictions of diclofenac for GUTS-scaled damage model.}
    \label{si:fig:model_fits_guts_scaled_damage_diclofenac}
\end{center}

\begin{center}
    \includegraphics[height=.27\textheight]{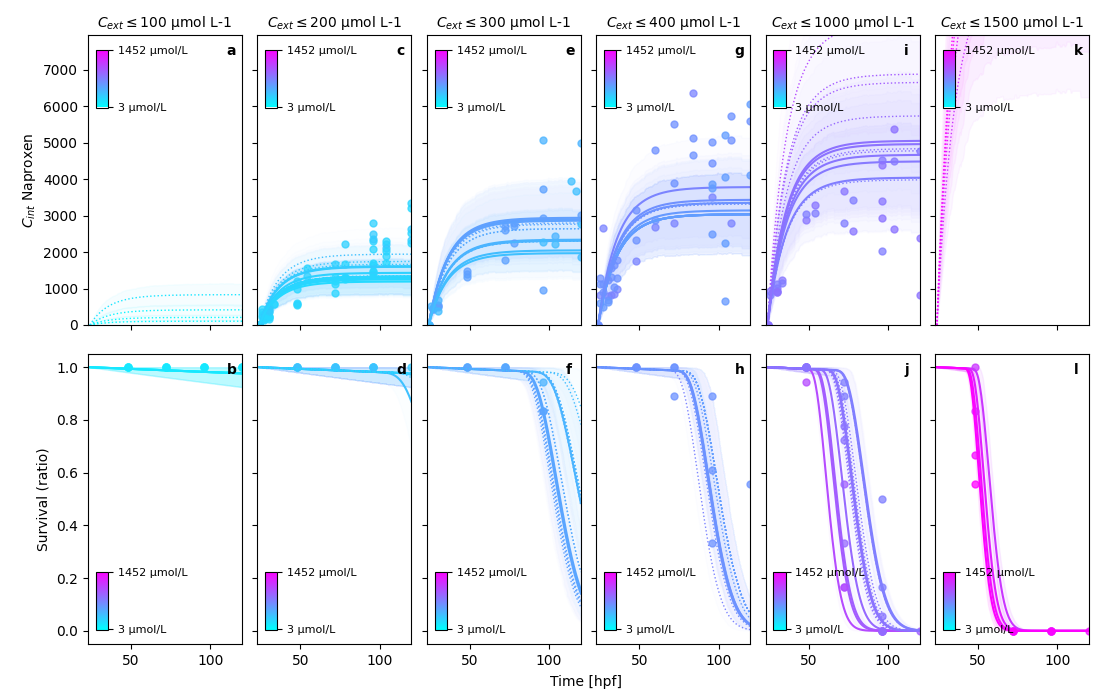}
    \captionof{figure}{Posterior predictions of naproxen for GUTS-scaled damage model.}
    \label{si:fig:model_fits_guts_scaled_damage_naproxen}
\end{center}

\begin{center}
    \centering
    \includegraphics[height=0.9\textheight]{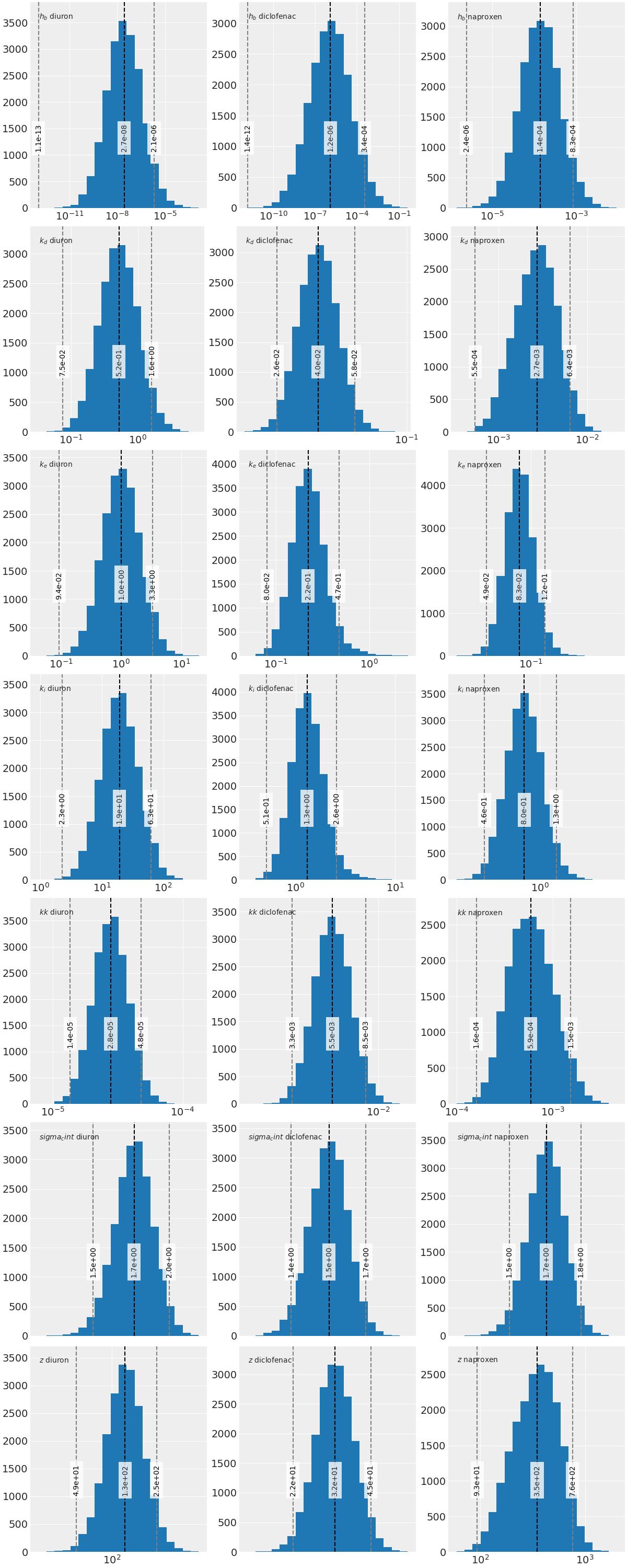}
    \captionof{figure}{Parameter estimates of the GUTS-scaled-damage model with substance specific parameters.}
    \label{si:fig:parameter_estimates_guts_scaled_damage}
\end{center}

\begin{minipage}\linewidth
\captionof{table}{Parameter estimates and posterior highest densitiy intervals (HDI) of the GUTS-scaled-damage model. The HDI contains 94\% of the probable parameter values given the data.}
\label{tab:parameters-guts_scaled_damage}
\begin{tabular}{llllllllll}
\toprule
Parameters & \multicolumn{3}{r}{Diuron} & \multicolumn{3}{r}{Diclofenac} & \multicolumn{3}{r}{Naproxen} \\
 & mean & hdi 3\% & hdi 97\% & mean & hdi 3\% & hdi 97\% & mean & hdi 3\% & hdi 97\% \\
\midrule
${k}_{i}$ & 25.01 & 2.94 & 59.77 & 1.44 & 0.53 & 2.49 & 0.83 & 0.46 & 1.23 \\
${k}_{e}$ & 1.30 & 0.11 & 3.13 & 0.25 & 0.08 & 0.45 & 0.09 & 0.05 & 0.12 \\
${k}_{d}$ & 0.66 & 0.08 & 1.52 & 0.04 & 0.03 & 0.06 & 0.00 & 0.00 & 0.01 \\
${z}$ & 139.63 & 52.23 & 241.07 & 32.67 & 21.99 & 44.61 & 384.24 & 90.52 & 726.16 \\
${k}_{k}$ & 0.00 & 0.00 & 0.00 & 0.01 & 0.00 & 0.01 & 0.00 & 0.00 & 0.00 \\
${h}_{b}$ & 0.00 & 0.00 & 0.00 & 0.00 & 0.00 & 0.00 & 0.00 & 0.00 & 0.00 \\
${\sigma}_{\text{cint}}$ & 1.73 & 1.50 & 1.97 & 1.52 & 1.39 & 1.64 & 1.65 & 1.49 & 1.81 \\
\bottomrule
\end{tabular}
\end{minipage}

\pagebreak
\subsection{GUTS-reduced model}

\subsubsection{Model description of the GUTS-reduced model}\label{si:mod:guts-reduced}

\begin{align}
\frac{dD}{dt} &= k_d \cdot (C_e-D) \\
h(t) &= k_k~ max(0, D(t) - z) +  h_b \\
S(t) &= e^{-\int_0^t h(t) dt}
\end{align}

\begin{table}[h!]
    \centering
    \footnotesize
        \caption{TKTD Parameters used in the GUTS-reduced model.}
        \label{si:tab:parameters-guts-reduced}
        \begin{tabularx}{\textwidth}{s X s}
        \toprule
        Parameter              & Definition & Unit\\
        \midrule
        ${k}_{d}$              & Dominant rate constant of damage dynamics & $h^{-1}$ \\
        ${z}$                  & Effect damage-threshold of the hazard function  & $\mu mol~L^{-1}$ \\
        ${k}_{k}$              & killing rate constant & $L~\mu mol^{-1}~h^{-1}$  \\
        ${h}_{b}$              & background hazard rate constant & $h^{-1}$ \\
        $\sigma_{\text{cint}}$ & Log-normal error of the internal concentration &  \\
        \bottomrule
        \end{tabularx}    
\end{table}

\pagebreak

\pagebreak
\subsubsection{Model fits for GUTS-reduced}\label{si:sec:model-fits-guts-reduced}

\begin{center}
    \includegraphics[height=.27\textheight]{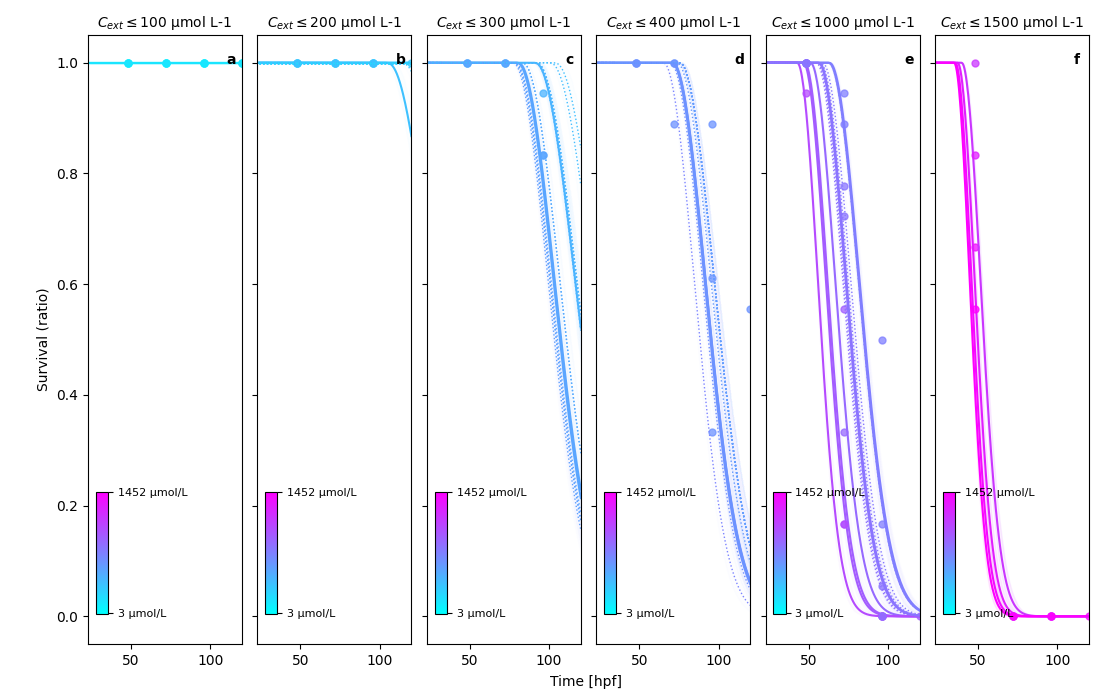}
    \captionof{figure}{Posterior predictions of naproxen for GUTS-reduced damage model.}
    \label{si:fig:model_fits_guts_reduced_naproxen}
\end{center}

\begin{center}
    \includegraphics[height=.27\textheight]{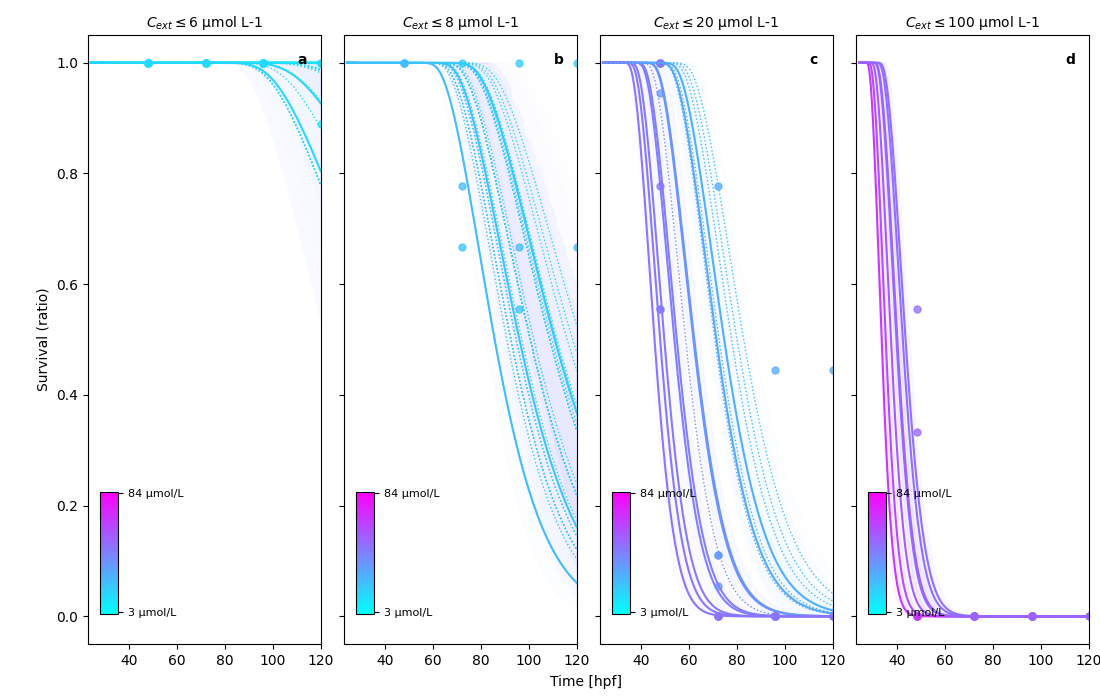}
    \captionof{figure}{Posterior predictions of diclofenac for GUTS-reduced damage model.}
    \label{si:fig:model_fits_guts_reduced_diclofenac}
\end{center}

\begin{center}
    \includegraphics[height=.27\textheight]{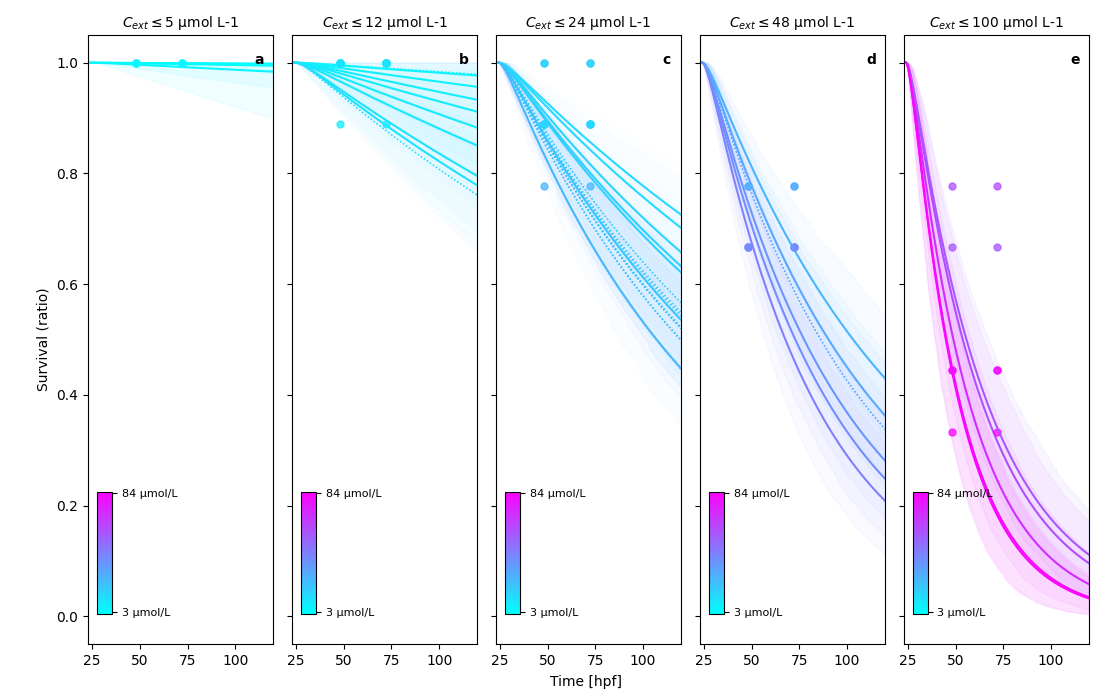}
    \captionof{figure}{Posterior predictions of diuron for GUTS-reduced damage model.}
    \label{si:fig:model_fits_guts_reduced_diuron}
\end{center}

\begin{center}
    \centering
    \includegraphics[height=0.9\textheight]{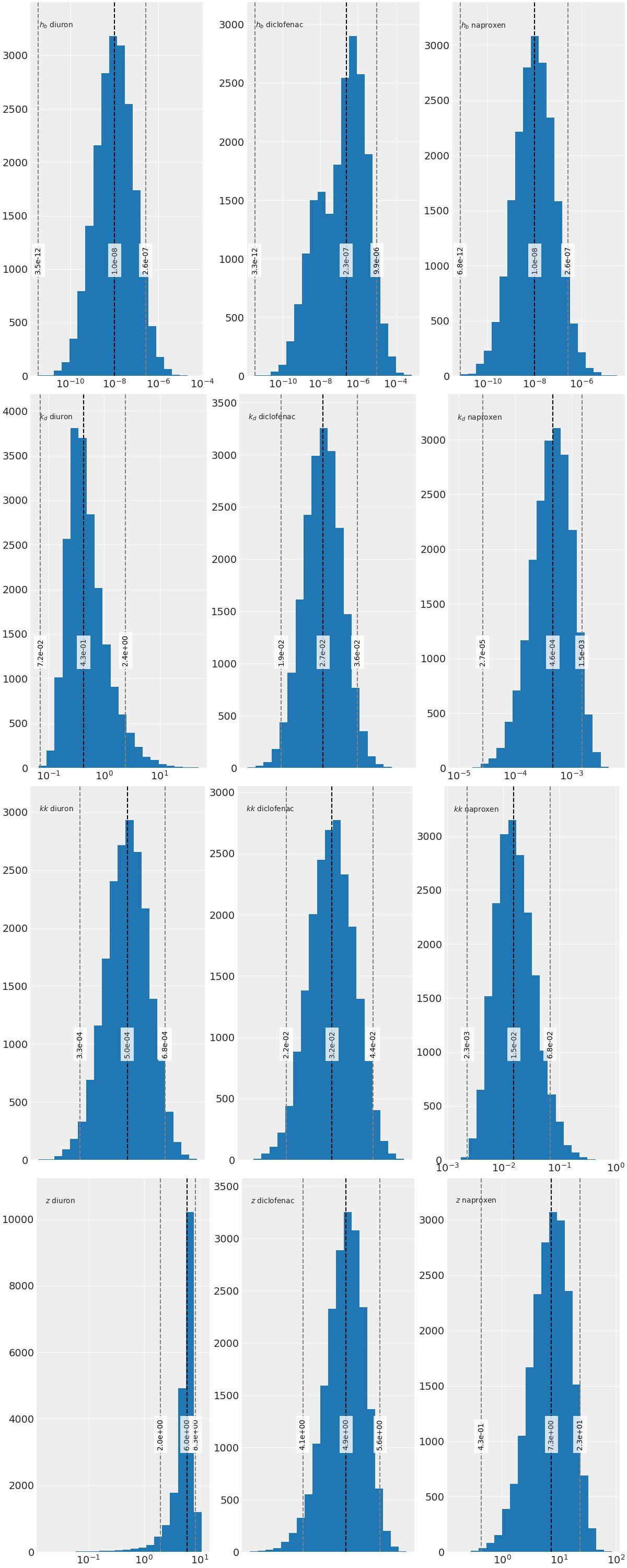}
    \captionof{figure}{Parameter estimates of the GUTS-reduced model with substance specific parameters.}
    \label{si:fig:parameter_estimates_guts_reduced}
\end{center}

\begin{minipage}\linewidth
\captionof{table}{Parameter estimates and posterior highest densitiy intervals (HDI) of the GUTS-reduced model. The HDI contains 94\% of the probable parameter values given the data.}
\label{tab:parameters-guts-reduced}
\begin{tabular}{llllllllll}
\toprule
Parameters & \multicolumn{3}{r}{Diuron} & \multicolumn{3}{r}{Diclofenac} & \multicolumn{3}{r}{Naproxen} \\
 & mean & hdi 3\% & hdi 97\% & mean & hdi 3\% & hdi 97\% & mean & hdi 3\% & hdi 97\% \\
\midrule
${k}_{d}$ & 0.80 & 0.08 & 2.17 & 0.03 & 0.02 & 0.04 & 0.00 & 0.00 & 0.00 \\
${z}$ & 5.60 & 2.20 & 8.35 & 4.85 & 4.11 & 5.59 & 9.24 & 0.43 & 22.17 \\
${k}_{k}$ & 0.00 & 0.00 & 0.00 & 0.03 & 0.02 & 0.04 & 0.02 & 0.00 & 0.06 \\
${h}_{b}$ & 0.00 & 0.00 & 0.00 & 0.00 & 0.00 & 0.00 & 0.00 & 0.00 & 0.00 \\
\bottomrule
\end{tabular}
\end{minipage}

\subsection{Estimated half-life of nrf2} \label{si:sec:half-life}

The half-life of \textit{nrf2} has been estimated at approximately 20 minutes \cite{Kobayashi.2004}. This value was compared against the posterior parameter distribution of the RNA-decay rate constant $k_{rd}$, inserting it in a simple exponential decay equation $R(t) = R_0^{- k_{rd}~t}$ and solving for $t$ at which $R(t) = R_0 / 2$, which gives the half-life of $t_{1/2} = \frac{ln(2)}{k_{rd}}$.

\begin{center}
    \includegraphics[width=.75\textwidth]{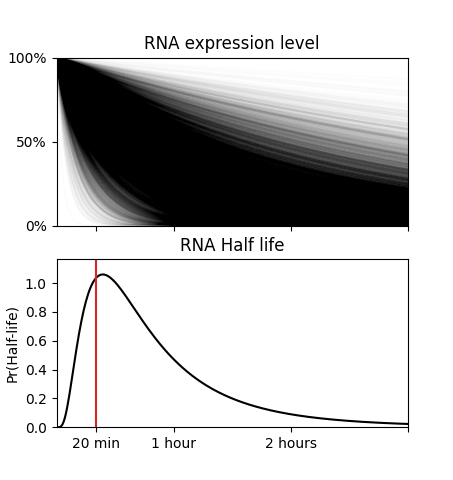}
    \captionof{figure}{Estimated half-life of RNA expression from the $r_{rd}$ parameter. A simple exponential decay model was assumed to estimate the half life. And a log-normal probability distribution was fitted to estimate the distribution of half-life times.}
    \label{si:fig:rna-half-life}
\end{center}

\subsection{Estimated half-life of proteins} \label{si:sec:half-life-protein}

The half-life of proteins is estimated to lie between 20--46 hours \cite{Harper.2016}. This value was compared against the posterior parameter distribution of the dominant rate constant for protein dynamics $k_{p}$, inserting it in a simple exponential decay equation $P(t) = P_0^{- k_{p}~t}$ and solving for $t$ at which $P(t) = P_0 / 2$, which gives the half-life of $t_{1/2} = \frac{ln(2)}{k_{p}}$.
Although the dominant rate constant lumps protein synthesis and decay together into one constant, it can give an idea of the approximate timescale of the dynamics. Seeing that the estimated half-life distribution matches the literature data to some degree. Can be seen as a confirmation that the approximate dynamics is captured correctly. Nevertheless, the true kinetics can only be estimated, once protein measurements are integrated into the model.

\begin{center}
    \includegraphics[width=.75\textwidth]{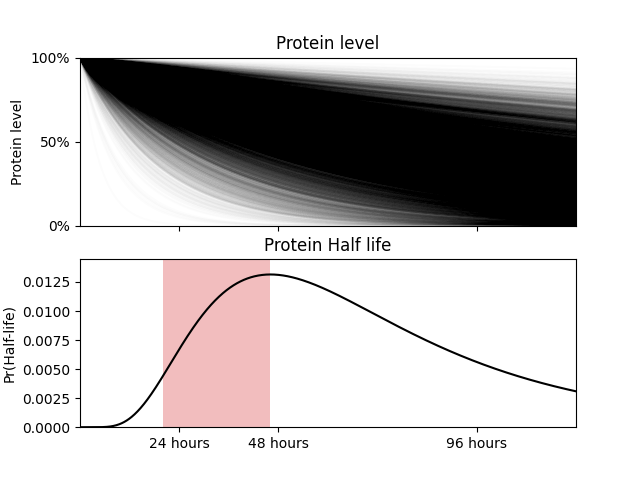}
    \captionof{figure}{Estimated half-life of proteins from the $r_{p}$ parameter. A simple exponential decay model was assumed to estimate the half-life. And a log-normal probability distribution was fitted to estimate the distribution of half-life times.}
    \label{si:fig:half-life-protein}
\end{center}

\end{document}